\documentclass[aps,prl,reprint,amsmath,amssymb,floatfix,longbibliography]{revtex4-2}

\usepackage{graphicx}
\usepackage{amsmath,amssymb,bbold,bm,color}
\usepackage{siunitx}
\usepackage{float}
\usepackage{epstopdf}
\usepackage{hyperref}
\usepackage[dvipsnames]{xcolor}
\usepackage[normalem]{ulem}

\usepackage[resetlabels,labeled]{multibib}

\newcommand{\bk}{{\bm k}}

\newcommand{\bp}{{\bm p}}

\newcommand{\bv}{{\bm v}}

\newcommand{\cT}{{\cal T}}
\newcommand{\cK}{{\cal K}}

\newcommand{\cH}{{\cal H}}

\newcommand{\cF}{{\cal F}}

\newcommand{\cV}{{\cal V}}

\newcommand{\bee}{\begin{equation}}
\newcommand{\ee}{\end{equation}}

\newcommand{\ket}[1]{| #1 \rangle}

\hypersetup{colorlinks=true,linkcolor=blue,citecolor=blue,urlcolor=blue}

\begin{document}

\title{High-temperature topological superconductivity in twisted
  double layer copper oxides}

\author{Oguzhan Can}
\author{Tarun Tummuru}
\author{Ryan P. Day}
\author{Ilya Elfimov}
\author{Andrea Damascelli}
\author{Marcel Franz}
\affiliation{Department of Physics and Astronomy, University of British Columbia, Vancouver BC, Canada V6T 1Z4} \affiliation{Quantum Matter Institute, University of British Columbia, Vancouver BC, Canada V6T 1Z4}

\date{\today}

%
\begin{abstract}
A great variety of novel phenomena occur when two-dimensional
materials, such as graphene or transition metal dichalcogenides, are
assembled into bilayers with a twist between individual layers. As a new
application of this paradigm, we consider 
structures composed of two monolayer-thin $d$-wave superconductors with
a twist angle $\theta$ that can
be realized by mechanically exfoliating van der Waals-bonded high-$T_c$ copper oxide
materials, such as Bi$_2$Sr$_2$CaCu$_2$O$_{8+\delta}$.
On the basis of symmetry arguments
and detailed microscopic modelling, we predict that for a range of twist
angles in the vicinity of $45^{\rm o}$, such bilayers form a robust, fully gapped
topological phase with spontaneously broken time-reversal
symmetry and protected chiral Majorana edge modes. When $\theta\approx 45^{\rm
  o}$, the topological phase
sets in at temperatures close to
the bulk $T_c\simeq 90$ K, thus furnishing  a long sought realization
of a true high-temperature topological superconductor.

\end{abstract}

\date{\today}
\maketitle


In a remarkable recent development Yu and co-workers \cite{Yuanbo2019} succeeded in
isolating and probing single monolayer samples of high-temperature
superconductor Bi$_2$Sr$_2$CaCu$_2$O$_{8+\delta}$ (Bi2212) with critical
temperatures $T_c$ essentially indistinguishable from the bulk $T_c\simeq 90$ K of
optimally doped  Bi2212 crystals. This result delivers the
long-awaited proof that high-$T_c$ superconductivity in Bi2212 (and possibly all the cuprates)
is an intrinsic property of a single two-dimensional monolayer. It also
opens exciting new possibilities for engineering structures with novel behaviour and functionality,
from individual high-$T_c$ cuprate monolayers.
Motivated by the results of Ref.\ \cite{Yuanbo2019} and recent
developments in van der Waals-bonded materials
\cite{Cao2018a,Cao2018b,Yankowitz2019,Sharpe2019,Efetov2019,Wang2019}, we consider here structures
composed of two cuprate monolayers with a twist angle $\theta$
between them.

 Twisted double layer graphene famously exhibits a sequence
of gate-voltage controlled correlated insulating and superconducting
phases \cite{Cao2018a,Cao2018b,Yankowitz2019,Sharpe2019,Efetov2019}  when $\theta$ is
close to the magic angle $1.1^{\rm o}$.  Similar physics is reported \cite{Wang2019} in twisted bilayer
WSe$_2$ in the range of angles between
$4.0^{\rm o}$ and $5.1^{\rm o}$. In both cases, the observed behaviour
is thought to reflect a subtle interplay between the band structure
energetics and topology brought about by the Moir\'{e}
patterns, electron-electron interactions and disorder \cite{Bistritzer2011,Nam2017,Kang2018,Zou2018,Po2018,Xie2018,Koshino2018,Guinea2018,Kang2019,Hejazi2019}.
As such, many aspects of the reported experimental observations
in twisted graphene and WSe$_2$
still await full theoretical understanding \cite{Andrei2020}. As we argue below, the
problem of twisted Bi2212 bilayers is potentially simpler to
analyze theoretically, at least at the level where each Bi2212
layer is treated as a BCS superconductor with a $d_{x^2-y^2}$ order
parameter. Results of such a theoretical analysis are, nevertheless,
non-trivial: over a continuous range of twist angles
$\theta$ in the vicinity of  $45^{\rm o}$, we find that below a critical temperature
$\tilde{T}_c(\theta)$ the bilayer system spontaneously breaks
time-reversal symmetry $\cT$ and forms a chiral topological phase that
can be thought of as an emergent $d_{x^2-y^2}\pm id_{xy}$
superconductor ($d\pm id'$ for short). Such a superconductor  has a fully gapped bulk and is characterized by
Chern number $C=\pm 2$ per layer with a gap of
$d_{x^2-y^2}\pm id_{xy}$ symmetry, respectively. It exhibits two gapless, topologically
protected  chiral  edge modes per monolayer. These modes are coherent
superpositions of electron and hole states and can be therefore
classified as Majorana particles \cite{Franz2015}. When  $\theta\approx 45^{\rm
  o}$, our results indicate that $\tilde{T}_c\simeq T_c$, suggesting that,
remarkably, in carefully assembled  Bi2212 samples, the
topological phase could set in well above the liquid nitrogen
temperature.

\begin{figure}[t]{
	\includegraphics[width = 6.5cm]{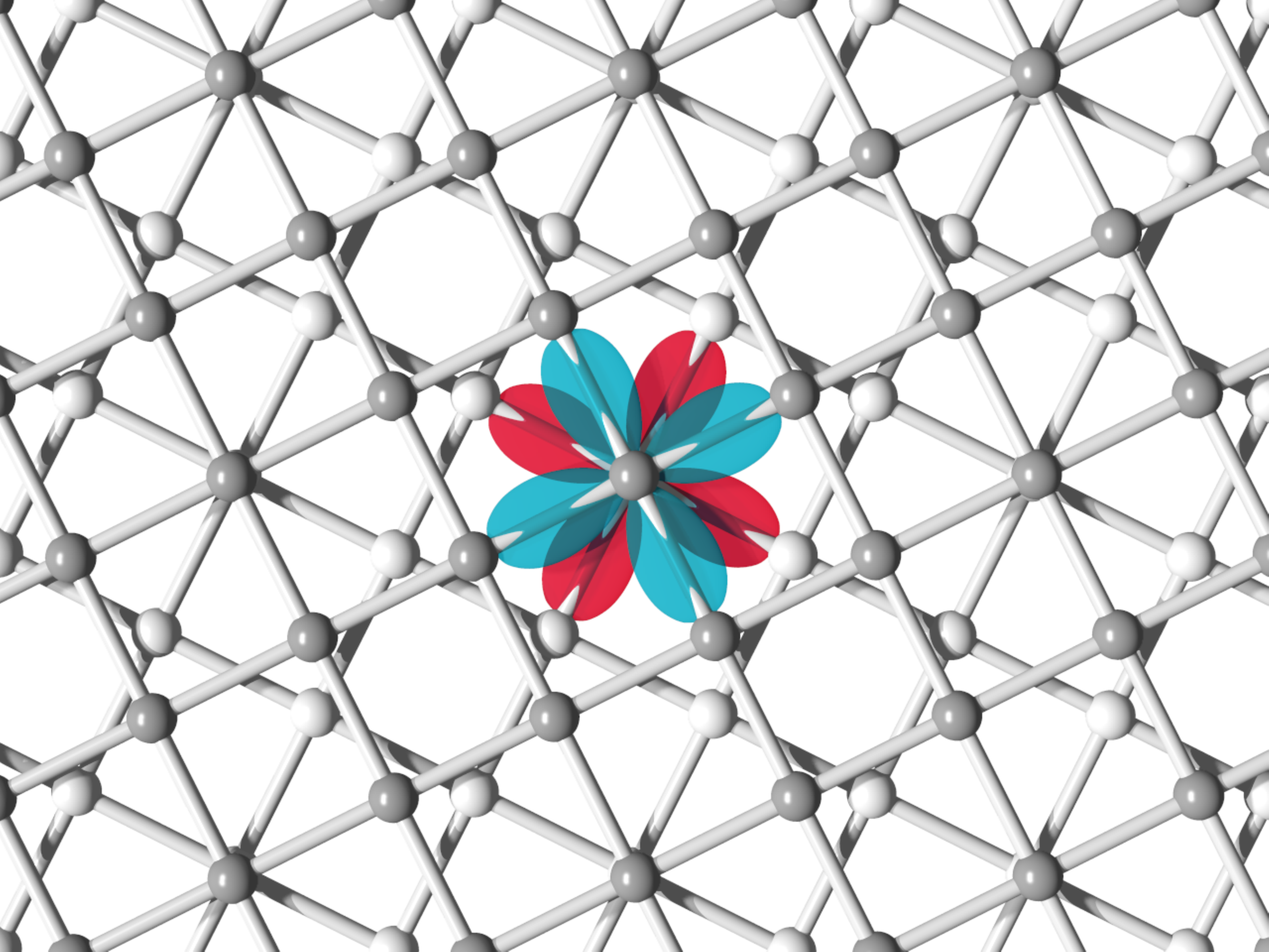}
	\caption{{\bf Schematic view of two copper-oxygen square
            lattices with twist angle close to $45^{\rm o}$.}
          Superconducting order parameter with a $d_{x^2-y^2}$ symmetry 
          in the twisted layer (red) resembles a $d_{xy}$ order parameter in
        the original untwisted coordinate frame (blue). Although each
        Bi2212 monolayer contains two CuO$_2$ planes, for simplicity
        and concreteness we primarily focus on models
        with a single CuO$_2$ plane per monolayer, as would be
        the case in Bi$_2$Sr$_2$CuO$_{6+\delta}$ and other high-$T_c$ compounds.}
\label{fig0}}
\end{figure}
The physics underlying the formation of the $d\pm id'$
topological phase is illustrated in Fig.\ \ref{fig0}. In momentum space, the $d_{x^2-y^2}$ order parameter that characterizes each
monolayer can be represented as
$\Delta_\bk=\Delta_0(\hat{k}_x^2-\hat{k}_y^2)$, where $\Delta_0$ denotes the amplitude
and $(\hat{k}_x,\hat{k}_y)=(k_x,k_y)/k$ are momentum components
defined relative to the principal crystal axes of the underlying
copper-oxygen lattice. When one of the layers is rotated by angle
$\theta$, its order parameter is transformed in the original coordinate frame to
\begin{equation}\label{e1}
\Delta_\bk^{(\theta)}=\Delta_0\left[\cos(2\theta)(\hat{k}_x^2-\hat{k}_y^2)+\sin(2\theta)
2\hat{k}_x\hat{k}_y\right].
\end{equation}
\begin{figure*}[t]{
	\includegraphics[width = 16.5cm]{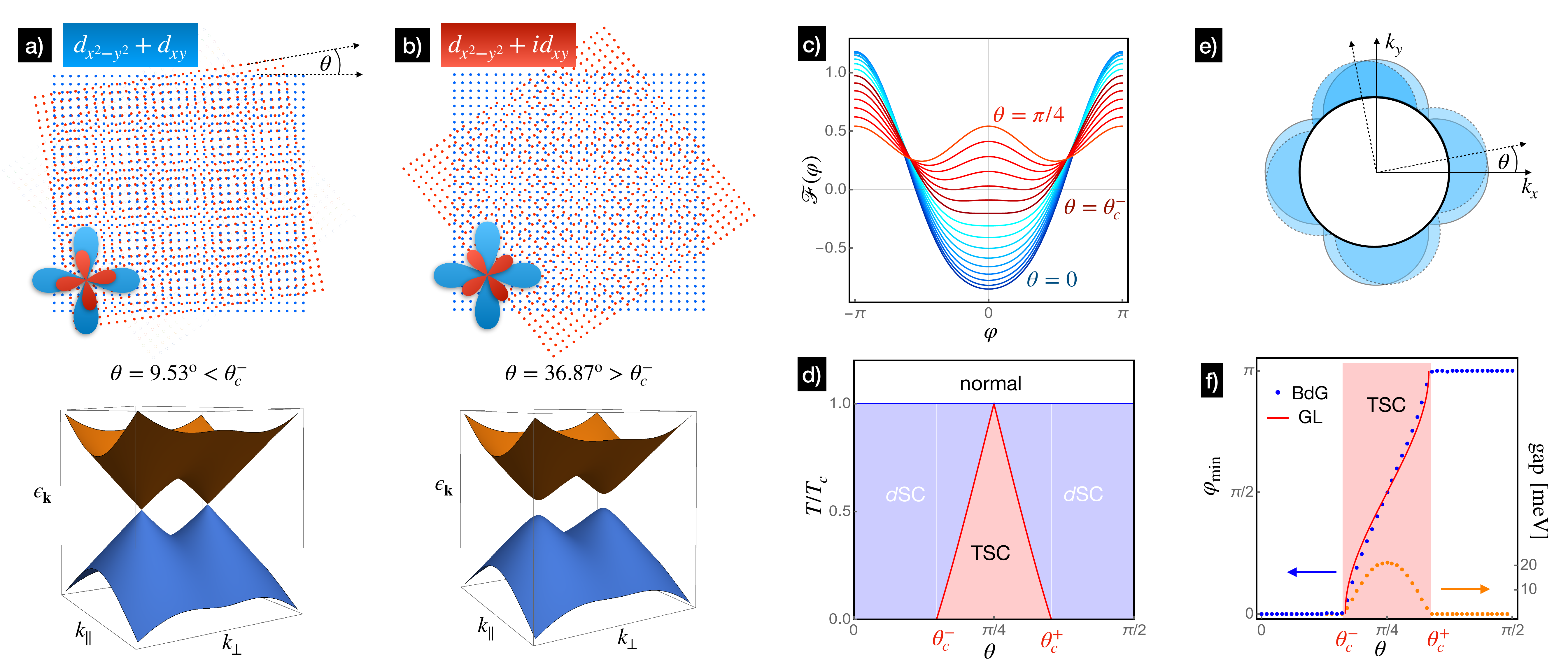}
	\caption{{\bf Twisted double layer $d$-wave superconductor.}
Panel (a) illustrates the lattice geometry at small twist angle
$\theta<\theta_c^-$ where the free energy is minimized
for interlayer phase difference $\varphi=0$. This results in a
gapless spectrum with weakly split Dirac points. For twist angles
$\theta>\theta_c^-$ in panel (b) the preferred state has $\varphi\neq
0$. It breaks time reversal and is fully gapped as indicated by the
massive Dirac spectrum. The spectra are obtained by diagonalizing
the BdG Hamiltonian \eqref{h4}; $k_\perp$ and $k_\parallel$ denote
momentum components  perpendicular and parallel to the Fermi surface
near a nodal point.
Panel (c) shows the GL free energy $\cF(\varphi)$ for twist angles $\theta$
  ranging from $0^{\rm o}$ to $45^{\rm o}$ in $3^{\rm o}$
  increments for $\cK=0.125$. When $\theta>\theta_c^-$ minima occur away from
  $\varphi=0$. Panel (d):  predicted phase diagram based on
  the the GL theory \eqref{e2}  and $\tilde{T}_c(\theta)$ given by
  Eq.\ \eqref{e6}.  TSC denotes the gapped
  topological phase, $d$SC stands for gapless $d$-wave superconductor.
Panel (e): Fermi surface and $d$-wave gap in the continuum
formulation of the microscopic model.
 Panel (f) displays the relative phase between the
  layers $\varphi_{\rm min}$ that minimizes the GL and the BdG free
  energy as a function of twist angle $\theta$.
  Orange symbols represent the
   spectral gap.}
	\label{fig1}}
\end{figure*}

From the vantage point of the unrotated monolayer, the second term in
Eq.\ \eqref{e1} has the functional form of a $d_{xy}$ order parameter. For coupled layers,
we then expect the Cooper pair tunneling to induce a subdominant $d_{xy}$ order
parameter in the unrotated layer. By symmetry, the same should happen in the other layer. Furthermore, if the two order
parameters combine with a complex phase, the result is a fully gapped
topological superconductor with broken $\cT$. The gapped nature of the phase
follows from the fact that the excitation gap is given by $|\Delta_\bk|$
which, for a  $d\pm id'$ gap function, is non-vanishing
everywhere at the Fermi
level, Fig.\ \ref{fig1}(a,b). $\cT$-breaking is evident because time reversal maps
$d+ id'$ to  $d- id'$. We demonstrate below that for a range of twist angles $\theta$ between
$\theta_c^-$ and $\theta_c^+$, such a $\cT$-broken phase becomes energetically
favorable below a critical temperature $\tilde{T}_c(\theta)$.

We note that $\cT$-broken  $d\pm id'$ and $d\pm is$ phases have been discussed
theoretically in various contexts, including bulk high-$T_c$ cuprates
in applied magnetic field \cite{Laughlin1998,Franz1998,Ashvin2001},
$c$-axis twist  Josephson junctions
\cite{Kuboki1996,Klemm2001,Tanaka2007,Hu2018}, and highly doped
single-layer graphene \cite{Baskaran2010,Levitov2012}. However,
conclusive experimental evidence for these states has never been
reported. A $\cT$-broken topological phase with $p_x\pm i p_y$ symmetry
has long been a leading candidate  for superconducting order in Sr$_2$RuO$_4$
\cite{Rice1995,Maeno1998,Kallin2009}. Recent Knight shift
measurements \cite{Pustogow2019}, however, appear to rule out this
possibility. By placing our emphasis on monolayer-thin samples of an established  $d_{x^2-y^2}$
superconductor, we argue that the present work promises to finally break the logjam in the quest for a chiral topological phase. As we show below, for  $\simeq 45^{\rm o}$ twist, the $\cT$-breaking  $d\pm id'$
phase  is nucleated robustly and reliably through the simple and well
understood mechanism of interlayer Cooper pair tunneling that is largely
insensitive to microscopic details.

\section{Ginzburg-Landau theory}
The phenomenon outlined above is most easily quantified by examining the
phenomenological Ginzburg-Landau (GL) theory. For the sake of
clarity and simplicity, we begin by considering models with a single
CuO$_2$ plane per monolayer. This is directly relevant to
Bi$_2$Sr$_2$CuO$_{6+\delta}$ (Bi2201), but, as we show in the subsequent
sections, our results are
broadly applicable to cuprates with multiple CuO$_2$ plane structure,
such as Bi2212, which has already been exfoliated \cite{Yuanbo2019} to the
monolayer limit. 

The GL free energy density for two coupled monolayers can be written as
\begin{eqnarray}
 \cF[\psi_1,\psi_2]&=&f_0[\psi_1]+f_0[\psi_2]+A |\psi_1|^2
                      |\psi_2|^2 \label{e2} \\
  &+&B(\psi_1\psi_2^\ast +{\rm c.c.})+C(\psi_1^2\psi_2^{\ast 2} +{\rm c.c.}),\nonumber
\end{eqnarray}
where $\psi_{a=1,2}$ are complex order parameters of the two layers and
$f_0[\psi]=\alpha|\psi|^2+{\frac{1}{2}}\beta|\psi|^4$ is the free energy
of a monolayer. We are interested in
spatially uniform solutions, so the gradient
terms have been omitted. If the two layers are physically identical, then the order parameter amplitudes
must be the same and the most general solution (up to an overall
phase) is
\begin{equation}\label{e3}
\psi_1=\psi, \ \ \ \psi_2=\psi e^{i\varphi},
\end{equation}
where we take $\psi$ to be real and positive.

Because $\psi_a$ describe order parameters with $d$-wave
symmetry, they transform as $\psi_a\to -\psi_a$ under a $90^{\rm o}$
rotation. Suppose we increase the twist angle $\theta$ continuously
from 0 to  $\pi/2$. Since $\psi_2$ changes sign under this operation, we
conclude that for the free energy \eqref{e2} to remain invariant, the
parameter $B$ must also change sign. The simplest dependence
consistent with this constraint is $B=-B_0\cos(2\theta)$, where we
take $B_0>0$. The negative sign is chosen to allow the phase difference between the layers to vanish as the twist angle approaches zero. This also
implies that $B=0$ for $\theta=\pi/4$, consistent with the intuition
that a $d_{x^2-y^2}$ Cooper pair cannot tunnel between the layers
in this configuration. Parameters $A$ and $C$ are not required by
symmetry to depend on $\theta$, so we treat them as constants henceforth. Incorporating these ingredients into Eq.\ \eqref{e2}, we find
\begin{equation}\label{e4}
\cF(\varphi)=\cF_0+2B_0\psi^2\left[-\cos(2\theta)\cos{\varphi}+ \cK\cos({2\varphi)}\right],
\end{equation}
where $\cK=C\psi^2/B_0$ and $\cF_0$ contains all terms independent of
$\varphi$. It is now a simple matter to
minimize  $\cF(\varphi)$ for a given twist angle $\theta$.  An
interesting scenario arises when $\cK>0$, as indicated by Fig.\
\ref{fig1}(c). Although symmetry alone does not
constrain the sign of $\cK$, a simple physical argument
  suggests that $\cK$ should, in general, be positive.  One can interpret the $\psi_1^2\psi_2^{\ast 2}$ term in Eq.\ \eqref{e2} as representing the
  coherent tunneling of {\em two} Cooper pairs between the layers. Up to an
overall  scale factor, its coefficient should, therefore, be
proportional  to the square of the coefficient  $B$, which itself corresponds to single pair
tunneling. Coefficient $C$ (and, hence, also $\cK$) will then be
positive, regardless of the sign of $B$.  Below, we demonstrate that microscopic 
models fully support this argument, yielding $\cK>0$.

At $\theta=45^{\rm o}$ and for $\cK>0$, the free energy minimum clearly
occurs at $\varphi_{\rm min}=\pm\pi/2$ which indicates a
$d\pm id'$ superconducting state.
For $\theta\neq 45^{\rm o}$, the free energy is minimized by
\begin{equation}\label{e5}
\varphi_{\rm min}=\arccos{\left(\frac{\cos{2\theta}}{ 4 \cK}\right)}.
\end{equation}
As illustrated in Fig.\ \ref{fig1}(f), a nontrivial phase difference occurs
for a range of angles
$\theta_c^-<\theta<\theta_c^+$ with $\theta_c^{\pm}=\frac{1}{2}\arccos{(\mp 4\cK)}$. This indicates a $\cT$-broken phase with order
parameter $d_{x^2-y^2}+e^{\pm i \varphi_{\rm min}} d_{xy}$ which is
fully gapped and topologically non-trivial as long as $\varphi_{\rm
  min}\neq 0,\pi$, see also Fig.\ \ref{fig1}(a,b). As $\cK$
depends on temperature through its dependence on $\psi$, the critical
temperature $\tilde{T}_c$ of the topological phase will be a function
of the twist angle $\theta$. If we adopt the standard GL
temperature dependence for the order parameter
$\psi(T)=\psi_0\sqrt{1-T/T_c}$, it is straightforward to deduce
\begin{equation}\label{e6}
\tilde{T}_c(\theta)=T_c\left(1-\frac{|\cos{2\theta}|}{ 4\cK_0}\right), \ \ \ \theta_c^-<\theta<\theta_c^+,
\end{equation}
where $\cK_0=C\psi_0^2/B_0$. This defines the phase diagram in
Fig.\ \ref{fig1}(d). We observe that, remarkably, $\tilde{T}_c(45^{\rm o})$
coincides with the $T_c$ of a single monolayer which can be as high as
90 K
in carefully prepared Bi2212 flakes. Away from the ideal $45^{\rm o}$
twist, the critical temperature falls approximately as
$\tilde{T}_c(\theta)\simeq T_c(1-|\theta-\pi/4|/2\cK_0)$.  Microscopic models
with realistic Bi2212 parameters discussed in the following give typical
values $\cK_0\simeq 0.1-0.2$, implying a significant extent for the
topological phase, as indicated in Fig.\ \ref{fig1}(d).

\section{Microscopic models}

We now turn to the microscopic theory. Although the pairing mechanism
in high-$T_c$ cuprates remains a subject of debate, it is widely accepted
that most physical properties of the superconducting state in the
optimally doped and overdoped regime are accurately described within
the framework of the standard  BCS theory with a $d_{x^2-y^2}$
order parameter. Therefore, we begin by modeling the twisted bilayer using a
simple continuum model of coupled $d$-wave superconductors, and complement
this with a calculation based on an attractive Hubbard model on the
square lattice. The continuum formulation has the advantage of being
applicable to an arbitrary twist angle $\theta$, similar to the
Bistritzer-MacDonald model for graphene \cite{Bistritzer2011}. In contrast with the continuum model above, the lattice calculation
captures various microscopic details of CuO$_2$ planes more effectively,
such as the shape of the Fermi surface. It is, however, limited to
commensurate twist angles that produce relatively small
Moir\'{e} unit cells due to practical constraints.

\begin{figure*}[t]{
    \includegraphics[width=\textwidth]{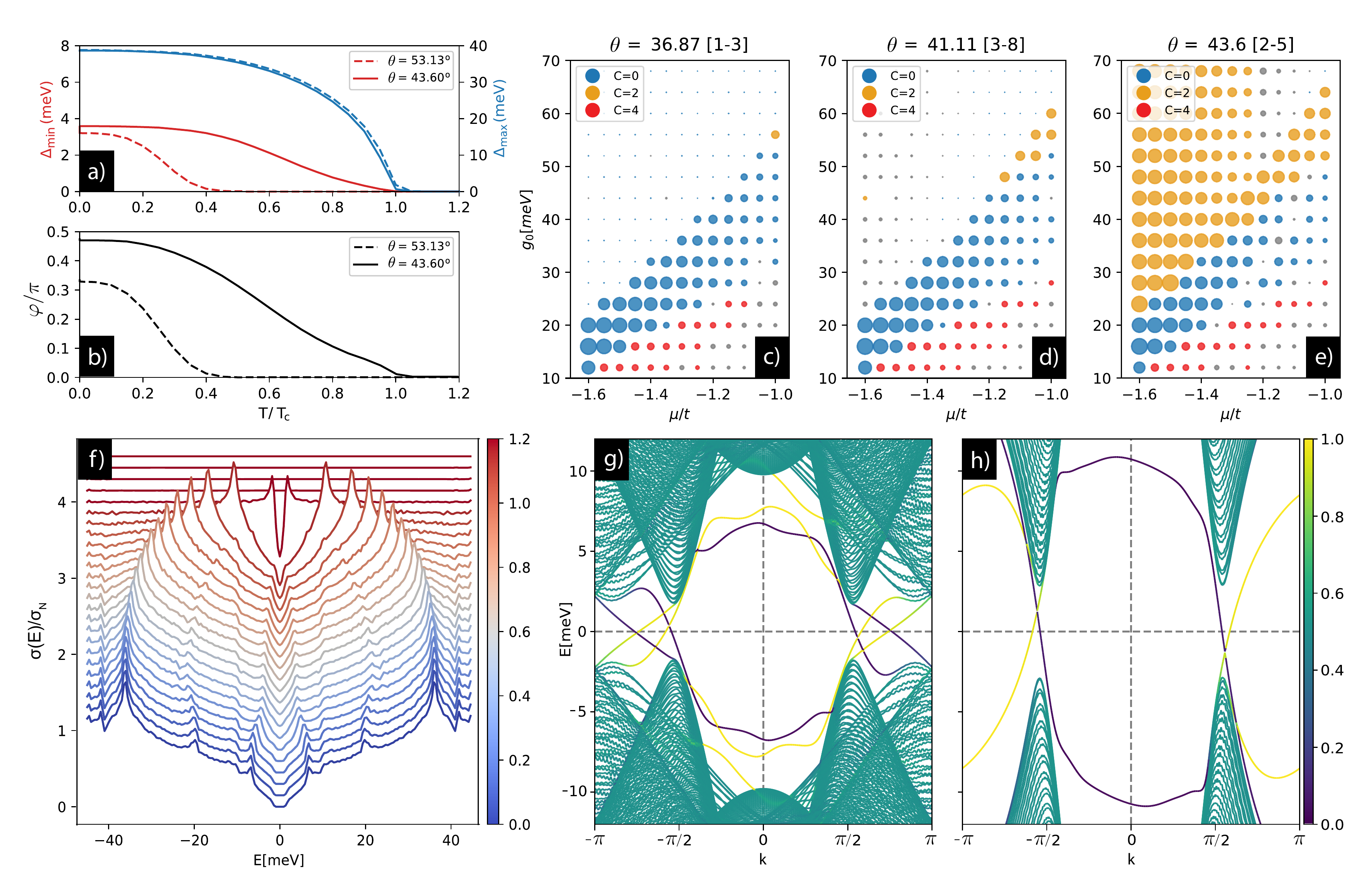}
	\caption{{\bf Lattice model results.}
Panels (a,b) show the temperature dependence of the minimum gap
$\Delta_{\rm min}$, the maximum gap $\Delta_{\rm max}$ and phase $\varphi$, based on a fully self-consistent lattice
calculation for coupled layers with commensurate  twist angles $\theta_{1,2}\simeq 53.13^{\rm  o}$,
and  $\theta_{2,5} \simeq 43.60^{\rm
  o}$ corresponding to a unit cell with 10 and 58 sites,
respectively. Panels (c-e) show zero temperature phase diagrams of the system for three
twist angles as a function of chemical potential $\mu$ and interlayer
coupling $g_0$. The range of chemical potentials $\mu \in
(-1.6t,-1.0t)$ correspond to range of fillings $n \in (0.033,0.04)$
near optimal doping. Each data point in the phase diagram corresponds
to an independent self-consistent solution and the radius of markers
is proportional to the size of the minimum gap  $\Delta_{\rm min}$. The
color indicates the Chern number as shown in the legend. Panel (f) displays
tunneling conductance $\sigma(E)/\sigma_N$ calculated for  $\theta\simeq 53.13^{\rm o}$, $\mu/t=-1.3$, $g_0=20$meV and temperatures ranging from 0 to
$1.2 T_c$ as indicated by color scale. Curves for different
temperatures have been offset for clarity. Panels (g) and (h) illustrate the edge
modes for $C=4$ and $C=2$ topological phases, respectively, in
$\theta_{2,5}$ configuration for parameters $\mu = -1.3 t$ and
$g_0=20, 52$ meV. The energy spectrum is shown for a bilayer system in the
infinite strip geometry with width of 90 unit cells. The color scale
represents the normalized position expectation value of the eigenstate along the
direction of finite length.
}
\label{fig2}}
\end{figure*}
In keeping with the strategy outlined above, we begin with a
single CuO$_2$ plane per monolayer, as for Bi2201. The continuum theory for such a
bilayer is defined by the Hamiltonian
\begin{eqnarray}\label{eq:continuumhamiltonian}
  \cH&=&\sum_{\bk \sigma a}\xi_{\bk a} c^\dag_{\bk\sigma a}c_{\bk\sigma a}
         +g\sum_{\bk \sigma}\left(c^\dag_{\bk\sigma 1}c_{\bk\sigma
         2}+{\rm h.c.}\right)   \label{h1} \\
  &+&\sum_{\bk a}\left(\Delta_{\bk a}c^\dag_{\bk\uparrow
      a}c^\dag_{-\bk\downarrow a}+{\rm h.c.}\right) - \sum_{\bk
      a}\Delta_{\bk a}\langle c^\dag_{\bk\uparrow
      a}c^\dag_{-\bk\downarrow a}\rangle. \nonumber
\end{eqnarray}
Here, $c^\dag_{\bk\sigma a}$ creates an electron with crystal momentum
$\bk$ and spin $\sigma$ in layer $a=1,2$ while $\xi_{\bk a}$ and $g$ represent
the in-plane kinetic energy and inter-plane tunneling amplitude,
respectively. The superconducting order parameter in layer $a$ can then be expressed as
\begin{equation}\label{h2}
\Delta_{\bk a}={\sum_\bp}' V_{\bk\bp}^{(a)}\langle c_{-\bp\downarrow
      a}c_{\bp\uparrow a}\rangle.
\end{equation}
The prime on
the summation indicates a restriction to momentum states with energy
within $\epsilon_c$ of the Fermi level. The
Hamiltonian \eqref{h1} should be regarded as a mean-field
approximation to the BCS pairing Hamiltonian with an interaction term
$\sum_{\bk\bp} V_{\bk\bp}^{(a)}  c^\dag_{\bk\uparrow
      a}c^\dag_{-\bk\downarrow a} c_{-\bp\downarrow
      a}c_{\bp\uparrow a}$, where $V_{\bk\bp}^{(a)}$ denotes the
    interaction matrix element in layer $a$. We shall use a simple
    separable form
\begin{equation}\label{h3}
V_{\bk\bp}^{(a)}=-2\cV \cos{(2\alpha_\bk)}\cos{(2\alpha_\bp)},
\end{equation}
where $\alpha_\bk$ represents the polar angle of the vector $\bk$. This is
known to yield a robust solution with $d_{x^2-y^2}$ symmetry
for a single CuO$_2$ layer, namely
$\Delta_{\bk a}=\Delta_d \cos{(2\alpha_\bk)}$.

In order to incorporate the twist, we take the interaction
in layer 1 as in Eq.\ \eqref{h3}, but we rotate the interaction
in layer 2 by angle $\theta$ such that $V_{\bk\bp}^{(2)}=-2\cV
\cos{(2\alpha_\bk-2\theta)}\cos{(2\alpha_\bp-2\theta)}$. For the sake
of simplicity, we consider a circular Fermi surface generated by
$\xi_{\bk a}=\hbar^2k^2/2m-\mu$ that remains invariant under rotation,
see Fig.\ \ref{fig1}(e). The
problem posed by Hamiltonian \eqref{h1} is then solved by defining
a four-component Nambu spinor $\Psi_\bk=(c_{\bk\uparrow
  1},c^\dag_{-\bk\downarrow 1}, c_{\bk\uparrow 2},c^\dag_{-\bk\downarrow
  2})^{T}$ in terms of which $\cH=\sum_\bk\Psi^\dag_\bk
h_\bk \Psi_\bk+E_0$.
Here, the Bogoliubov-de Gennes (BdG) Hamiltonian reads
\begin{equation}\label{h4}
  h_\bk=
 \begin{pmatrix}
   \xi_\bk & \Delta_{\bk 1} & g & 0 \\
   \Delta_{\bk 1} ^\ast & -\xi_\bk & 0 & -g \\
   g & 0 &   \xi_\bk & \Delta_{\bk 2} \\
   0 & -g &    \Delta_{\bk 2} ^\ast & -\xi_\bk
  \end{pmatrix}
\end{equation}
and $E_0=\sum_\bk 2\xi_\bk -\sum_{\bk  a}\Delta_{\bk a}\langle
c^\dag_{\bk\uparrow a}c^\dag_{-\bk\downarrow a}\rangle$. Diagonalizing
$h_\bk$ gives two pairs of energy eigenvalues $\pm E_{\bk \alpha}$
$(\alpha=1,2$) for each momentum $\bk$. With the
Hamiltonian \eqref{h1} expressed in diagonal form, the free energy of the
system can be calculated from the standard expression
\begin{equation}\label{h5}
 \cF_{\rm BdG}=E_0-2 k_BT\sum_{\bk \alpha}\ln\left[2\cosh{(E_{\bk \alpha}/2 k_BT)}\right].
\end{equation}
By performing a systematic expansion of $\cF_{\rm BdG}$ in powers of the
  order parameter amplitudes $\Delta_d$, it is possible to ascertain
  various coefficients entering the GL free energy \eqref{e2}. This
  confirms the form of the
   GL coefficients $B$ and $C$ deduced
previously on the basis of general symmetry
  arguments, and that $\cK$ in Eq.\ \eqref{e4} is positive.  This
  calculation is summarized in Methods.

We proceed with a fully self-consistent solution which follows from
minimizing $\cF_{\rm BdG}$ with respect to $\Delta_{\bk a}$ and can be
performed for any
given $\cV$, twist angle $\theta$ and temperature $T$. To facilitate
this calculation we assume that the order parameters in the two layers have
the same amplitude $\Delta_d$, but can differ in phase as in Eq.\
\eqref{e3}. Free energy  computed from Eq.\ \eqref{h5} then shows the same
qualitative behaviour as the GL  theory \eqref{e4} with $\cK >0$: for small twist
angles, $\cF_{\rm BdG}(\varphi)$ has
a single minimum at $\varphi=0$, while for twist angles close to
$45^{\rm o}$ there are always two degenerate minima at $\varphi=\pm\varphi_{\rm
  min}$, indicating formation of a $\cT$-broken topological
phase. Fig.\ \ref{fig1}(f) shows the calculated $\varphi_{\rm  min}$
as a function of the twist angle $\theta$ for realistic Bi2212
parameters \cite{andersen1995lda,Dama2003,Fischer2007} $\Delta_d=40$
  meV, $g=30$ meV, $\epsilon_c=60$ meV at $T=0$. We observe an excellent
  agreement with the prediction of the GL theory when $\cK=0.125$. The
  BdG theory also allows us to predict the minimum excitation gap
  which, in this simple model, can be as large as 20 meV when $\theta$ is close to  $45^{\rm  o}$.

The BdG theory can be formulated directly on the lattice by starting
from the Hubbard model with on-site repulsion and nearest-neighbour
attraction. This is known to produce a $d_{x^2-y^2}$ superconductor
when applied to a single CuO$_2$ monolayer.  The Hamiltonian is
\begin{eqnarray}
  H&=& - \sum_{ ij, \sigma a} t_{ij}c^\dag_{i \sigma a} c_{j \sigma a} 
  - \mu \sum_{i \sigma a} n_{i \sigma a} \nonumber \\
  &+&\sum_{ ij,a}V_{ij}n_{ia}n_{ja}
- \sum_{i j \sigma} g_{ij} c^\dag_{i \sigma 1} c_{j \sigma 2},
    \label{hm_latt}
\end{eqnarray}
where $t_{ij}$ encodes the normal-state band structure of the single
layer, $g_{ij}$ describes the interlayer tunneling and $V_{ij}$ denotes
density-density interactions.  The mean-field calculations are performed
in bilayer geometries characterized by a twist vector $\bv=(m,n)$ and  commensurate twist angle $\theta_{m,n}=2\arctan{(m/n)}$ as
explained in Methods. Fig.\ \ref{fig2}
summarizes our main results. These lend
further support to our conclusions drawn on the basis of the GL and
continuum BdG approaches and show additional interesting features.

 The lattice model confirms the
onset of the $\cT$-breaking phase below critical temperature
$\tilde{T}_c(\theta)$ whose dependence on $\theta$ is consistent with
the GL prediction, see Fig.\ \ref{fig2}(a,b). The results are in reasonable quantitative agreement with the
continuum BdG theory but the lattice model gives a smaller
spectral gap.  This may be attributed to more complicated Fermi surface
geometry resulting from Brillouin zone folding that accompanies the large
real-space Moir\'e unit cell. The lattice model allows for a direct
evaluation of the Chern number $C$, as discussed in Methods. 
In addition to the $C=4$ topological
phase anticipated on the basis of the continuum BdG theory, the $T=0$ phase
diagrams in Fig.\ \ref{fig2}(c-e) reveal the
existence of $C=2$ and $C=0$ gapped phases that can be reached by
varying the chemical potential $\mu$ and the interlayer coupling
strength $g_0$. In an experiment, the former can be tuned over a wide range by oxygen annealing
\cite{Yuanbo2019}, while the latter may depend on the twist angle
$\theta$ and could be further controlled by applying hydrostatic pressure as demonstrated in twisted graphene \cite{Yankowitz2019}. As explained in Methods, the $C=0$ phase corresponds to a
$d+is$ superconductor, while the $C=2$ phase can be effectively thought of 
as a single-layer $d+id'$ superconductor. For a strip geometry, the lattice model
predicts protected chiral edge modes traversing the bulk
gap, Fig.\ \ref{fig2}(g,h), confirming the topological nature of
the $d+id'$ phase.

Additional interesting behaviour is observed at nonzero temperature. Fig.\ \ref{fig:finiteT-phasediag} in Methods shows a phase transition from $d+is$ to $d+id$ phase driven by increasing $T$. As we discuss in more detail below, all these phenomena are experimentally accessible using standard spectroscopic, thermodynamic and transport techniques.

\section{Stability analysis and the Density functional theory}

In order to establish the physical stability of the twisted double layer
structure and to estimate the interlayer coupling, we performed extensive Density Functional Theory (DFT)
simulations. We have employed the structure of Bi2201; with a single CuO$_2$ plane per monolayer, it has a direct connection to the $N=1$ layer continuum and lattice models. Furthermore, the volume of the unit cell is greatly reduced with respect to Bi2212, facilitating calculation of the freestanding bilayer heterostructure. This analysis is focused on the coupling between two CuO$_2$ blocks; as for the previous section, we do not expect the additional internal structure of each block in the $N>1$ case to change our results substantively. 

\begin{figure*}[t]{
	\includegraphics[width = \textwidth]{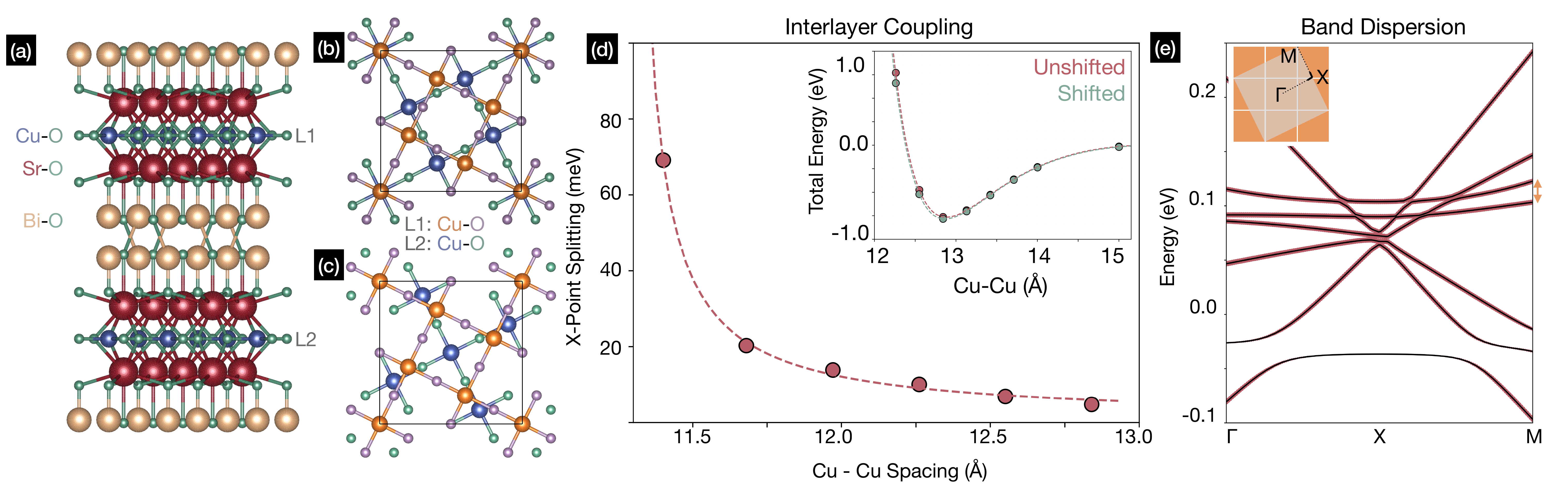}
	\caption{{\bf Crystal structure and results of the twisted double layer DFT calculations.} In panel (a), side view of the unit cell with unshifted monolayers. Top view of the CuO$_2$ planes are illustrated for the unshifted (b) and shifted (c) unit cells. For the unit cell in (b), we plot the splitting at the $X$ point as a function of the Cu - Cu distance in (d). A narrow window of the twisted bilayer bandstructure near the $X$ point of the original Bi2201 Brillouin zone is provided in (e), at a Cu-Cu spacing of 11.97 \AA . The pair of bands used for defining the splitting in (d) are identified by the orange arrow. A fit of the $X$-point splitting to $1/x$ is shown as a dashed red line in (d). The equilibrium layer spacing is solved for as the minimum in the system's total energy, as illustrated by the inset. The total energy of the bilayer unit cell is given for both unit cells in (b,c). The energy scale is defined relative to the limit of two isolated monolayers. 
} 
	\label{fig3}}
\end{figure*} 
 Bi2201 supercells in the $\bv=(1,2)$ geometry were constructed from a
 bulk model defined with an in-plane Cu-Cu distance of $a = 3.99$ \AA, and unit-cell height of $c = 23.98$ \AA . The tetragonal
 $I4/mmm$ space group was imposed to minimize complications due to the
 small orthorhombicity of the actual material. Two unit cells, each
 containing a single CuO$_2$ plane and counter-rotated by the twist
 angle were combined in a freestanding bilayer construction, with a
 vacuum buffer of 15\AA,\  see Fig. \ref{fig3}(a). We considered two
 high-symmetry configurations of the bilayer unit cell, with the two
 layers translated by (0,0) and ($\sqrt{5}a/2$,$\sqrt{5}a/2$), as
 illustrated in Fig.\ \ref{fig3}(b,c).The Kohn-Sham equations were
 solved self-consistently using the SCAN meta-GGA exchange correlation potential \cite{Sun_2015}, as implemented in VASP \cite{Kresse_1993,Kresse_1996}. As opposed to standard GGA, the recently developed SCAN functional incorporates kinetic-energy corrections, supporting the study of mixed-bonding environments \cite{Sun_2015}.  In bilayer cuprates, weak van der Waals and ionic bonding coexist, making this approach well-suited to the problem \cite{Furness_2018, Zhang_2020}. This circumvents the need for empirical van der Waals corrections to the GGA. We repeated this process for various
 inter-bilayer distances to ascertain the energetic stability of the
 bilayer compound. In the inset of panel (d), we plot the total ground
 state energy, evaluated relative to the limit of two independent
 monolayers. In both translational configurations, we find a similar
 energy minimum near 12.87 \AA.  We note that this is the distance
 between neighbouring CuO$_2$ layers; the BiO layers which constitute
 the bilayer interface are 9.45 \AA closer together. These results are largely independent of the exchange correlation potential used, as explored in Methods.

 To estimate the strength of interlayer coupling between neighbouring CuO$_2$ layers, we identify an avoided crossing at the $X$ point of the Bi2201 Brillouin zone, indicated in the bandstructure of Fig. \ref{fig3}(e). The size of this splitting decays with the Cu-Cu distance, to a value of $\sim$ 5 meV at the optimal spacing of 12.87 \AA, as shown in Fig. \ref{fig3}(d). We emphasize that this splitting does not provide a direct measure of the parameter $g_0$, but establishes evidence for finite Cu-Cu interlayer coupling within the energy scale surveyed in the text. It is important to note that there is precedent for a renormalization of the DFT-derived interlayer coupling strength in cuprates, as measured via photoemission. These renormalization factors vary over a range of at least 2-5 in different materials \cite{Markiewicz2005}. Additional details of the calculations presented here can be found in the Methods section.

\section{Outlook}

Our results establish a new avenue for engineering  high-temperature
topological superconductors by combining monolayers of
van der Waals-bonded  $d$-wave superconducting cuprates, such as
Bi$_2$Sr$_2$CuO$_{6+\delta}$ or
Bi$_2$Sr$_2$CaCu$_2$O$_{8+\delta}$ \cite{Yuanbo2019}. The topological
phase is fully gapped, breaks time reversal symmetry and supports
protected chiral  Majorana modes at the sample boundary. 
The chiral $d\pm id'$ phase arises robustly, driven by
Cooper pair tunneling between the layers when they are twisted by
an angle close to 45$^{\rm o}$. Unlike in graphene, where the twist
angle must be tuned very accurately, we predict that a wide range of
angles (45$^{\rm o}\pm 10^{\rm o}$) will produce nontrivial physics in
these systems. 

The resulting fully gapped
topological phase can  be probed by conventional spectroscopic or
transport techniques. For example, on lowering the sample temperature,
angle resolved photoemission
or tunneling spectroscopies \cite{Dama2003,Fischer2007} will detect the transition at
$T=\tilde{T}_c(\theta)$ from the gapless $d_{x^2-y^2}$ superconductor
with Dirac nodes to a fully gapped excitation spectrum. This
change should also be observable via thermodynamic probes such as those
measuring the specific heat \cite{Riggs2011} and superfluid density \cite{Tallon2003,Hosseini2004}.
Signatures of the broken $\cT$-symmetry characteristic of the $d\pm id'$ and $d+is$ phases
may be identified by circular birefringence and dichroism
experiments \cite{Yip1992}. 
In transport, the non-zero Chern number manifests as quantized
electronic contribution to the thermal Hall conductance $\kappa_{xy}=
C(\pi^2 k_B^2 T/3 h)$ \cite{Ashvin2001}. In mesoscopic samples,
longitudinal thermal conductance $\kappa_{xx}$ will also
be quantized due to the ballistic contribution of the edge states.
The $d\pm id'$ phase exhibits anomalous interlayer
Josephson effect due to the two-minima structure of the the free energy depicted
in Fig.\ \ref{fig1}(c); the Josephson current
$I_J= (2e/\hbar)\partial\cF/\partial\varphi$ will develop pronounced
deviations from the usual sinusoidal dependence on the phase bias
$\Delta\varphi=\varphi-\varphi_{\rm min}$.

It is worth noting that existing experimental results on $c$-axis
twist Josephson tunneling in cuprates already offer some
support for the ideas advanced here. According to symmetry analysis,
which assumes unbroken time reversal and pure $d_{x^2-y^2}$ order
parameter, the Josephson current $I_J$ must vanish when $\theta=45^{\rm
  o}$ \cite{Kuboki1996,Klemm2001,Tanaka2007}. In Methods, we show that for a $d+id'$ state  $I_J$, however, remains 
nonzero at $\theta=45^{\rm o}$. Non-vanishing $I_J$ has often been
interpreted as evidence for conventional $s$-wave order
parameter in experiments \cite{LI19971495,zhu2019isotropic}. Given what we know about the order parameter symmetry in cuprates
today, one may argue that a more plausible explanation for
non-vanishing $I_J$ is the spontaneously $\cT$-broken phase advocated
in this work. 

The authors are indebted to D.A. Bonn, D.M.  Broun, J.A. Folk, C. Kallin, C. Li, \'E. Lantagne-Hurtubise,
S. Plugge, S. Sahoo, O. Vafek and Z. Ye for valuable discussions and
correspondence.  The work described here was supported by NSERC.
This research was undertaken thanks in part to funding from the Max Planck-UBC-UTokyo Centre for Quantum Materials and the Canada First Research Excellence Fund, Quantum Materials and Future Technologies Program. OC is supported by International Doctoral Fellowship from UBC.

\subsection{Data availability}
This manuscript contains no experimental data.

\subsection{Code availability}
The complete code used to obtain results shown in Fig. \ref{fig2} will
be made publicly available at https://github.com/ocanphys/tbcuprate/. DFT results shown
in Fig.\  \ref{fig3} were obtained VASP \cite{Kresse_1993,Kresse_1996} . Input files are also available through our repository.

 \bibliography{did}


\newpage

\pagebreak

\section{Methods}

\renewcommand{\bibnumfmt}[1]{[S#1]}
\renewcommand{\citenumfont}[1]{S#1}

\subsection{Microscopic derivation of GL coefficients}
From the continuum BdG model, defined by the matrix Hamiltonian
\eqref{h4}, it is possible to derive  the corresponding GL theory by
directly expanding the free energy $\cF_{\rm BdG}$ given in Eq.\ \eqref{h5} in
the powers of the superconducting order parameters.
This procedure is made feasible by the fact that explicit expressions
for the  two positive-energy eigenvalues of $h_\bk$ can be obtained
and are given by
\begin{equation}
  E_{\bk\alpha}=\sqrt{(\Delta_{\bk 1}^2+\Delta_{\bk 2}^2)/2
                   +\xi_\bk^2+g^2+(-)^\alpha D_\bk}
    \label{gl1}
\end{equation}
where $D_k^2=(\Delta_{\bk 1}^2-\Delta_{\bk 2}^2)^2/4+g^2(\Delta_{\bk
  1}^2+\Delta_{\bk 2}^2+4\xi_\bk^2)-2g^2 \Delta_{\bk 1}\Delta_{\bk
  2}\cos\varphi$ and $\alpha=1,2$. We now express the two $d$-wave
gap functions as 
\begin{equation}   \label{gl2}
 \Delta_{\bk 1}=\psi\cos(2\alpha_\bk), \ \ \ \Delta_{\bk 2}=\psi\cos(2\alpha_\bk-2\theta),
\end{equation}
substitute into \eqref{gl1} and expand the free energy \eqref{h5} in
the powers of the order parameter amplitude $\psi$ to fourth order. This
expansions results in many terms and is best carried out with the help of
Mathematica or similar software capable of symbolic manipulations.

\begin{figure}[t]
    \includegraphics[width=1.\columnwidth]{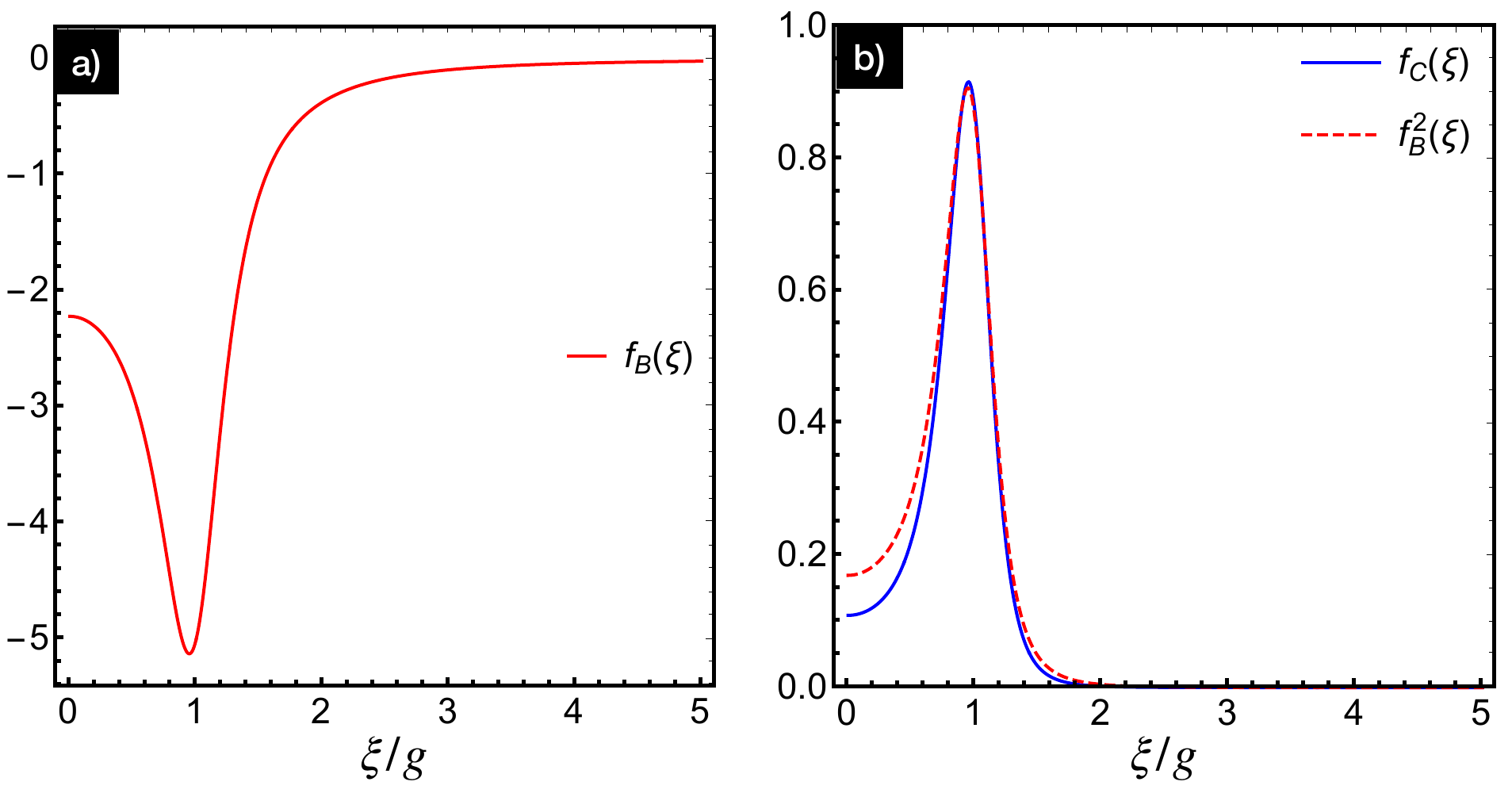}
    \caption{{\bf Analysis of the energy integrands.} (a) Integrand
      $f_B(\xi)$ entering the expression Eq.\ \eqref{gl3} for the GL
      theory coefficient $B$. (b) Integrand $f_C(\xi)$ entering the
      expression for the GL coefficient $C$ compared to $f_B(\xi)^2$. Note
    that the amplitude of the latter is scaled by a constant factor of
  1/30 for easier comparison. The temperature is chosen such that
  $k_BT=0.2g$ but the qualitative properties of the integrands remain
  the same in a wide range of temperatures.}
  \label{glc}
\end{figure}

We are primarily interested in terms governing the phase difference
between the two layers, that is, coefficients $B$ and $C$ in the GL
free energy \eqref{e2}. These can be isolated by focusing on
terms containing powers of $\cos\varphi$. Coefficient $B$ is thus
found as a prefactor of $\psi^2 \cos{\varphi}$ in the above expansion
and reads
\begin{equation}   \label{gl3}
B=\cos{(2\theta)}N_F\sum_{s=\pm}\int_0^{\epsilon_c}d\xi \frac{-\pi
  sg}{\xi |\xi-sg|}\frac{\tanh{|\xi-sg|}}{ 2k_BT} 
\end{equation}
In deriving this expression we replaced the crystal momentum sum using
the usual prescription $\sum_\bk\to N_F\int_{-\epsilon_c}^{\epsilon_c}
d\xi\int_0^{2\pi}d\phi$ where $N_F=L^2m/\hbar^2$ is the normal-state
density of states at the Fermi level, $L$ is the system size and the
angular integral has been performed. The remaining integral over the
energy variable cannot be completed in the closed form but one can show that
the integrand $f_B(\xi)$ is {\em negative} for all values of $\xi$ and thus the
integral itself is also negative, as expected on physical
grounds. This is illustrated in Fig.\ \ref{glc}(a).

Coefficient $C$ can likewise be identified as a prefactor of the $\psi^4
\cos{2\varphi}$ term in the expansion and has the form $C=(2+\cos{4\theta}) I_C$. Here $I_C=N_F\int_0^{\epsilon_c}f_C(\xi)d\xi $ is the corresponding energy integral similar to Eq.\ \eqref{gl3}. Its integrand $f_C(\xi)$ has a more complicated structure and for the sake of brevity we do not show it
explicitly. Importantly $f_C(\xi)$ is {\em positive} for all $\xi$
and, as anticipated in the main text, is approximately equal to the
integrand of Eq.\ \eqref{gl3} squared, as shown in Fig.\ \ref{glc}(b). 
We thus conclude that the continuum BdG model indeed supports the GL
phenomenology deduced in the main text on the basis of general physical
arguments. Specifically the calculation explicitly shows that  prefactor $\cK$
of the $\cos{2\varphi}$ term in Eq.\ \eqref{e4} is positive which we
argued was
the key condition underlying the emergence of the $\cT$-broken
topological phase in the twisted $d$SC bilayers.

\subsection{Self-consistent mean field theory on the lattice}

To describe the two-layer system microscopically, we employ the
Hubbard Hamiltonian \eqref{hm_latt} and treat the interaction term at
the mean field level to account for 
$d$-wave superconductivity in each layer.  The MF Hamiltonian becomes
\begin{eqnarray}
  H&=& -t \sum_{\langle ij \rangle \sigma a} c^\dag_{i \sigma a} c_{j \sigma a} 
  -t' \sum_{\langle\langle ij \rangle\rangle \sigma a} c^\dag_{i \sigma a} c_{j \sigma a}  
  - \mu \sum_{i \sigma a} n_{i \sigma a} \nonumber \\
  &+&\sum_{\langle ij \rangle a}\left(\Delta_{ij, a}c^\dag_{i\uparrow
      a}c^\dag_{j\downarrow a}+{\rm h.c.}\right) 
- \sum_{i j \sigma} g_{ij} c^\dag_{i \sigma 1} c_{j \sigma 2}, 
    \label{hm_latt1}
\end{eqnarray}
where  $t$ and $t'$ are the hopping amplitudes between first and second
neighbour sites on the square lattice, $\mu$ is the chemical potential
that controls on-site particle density $n_{i \sigma a}$ and
$\Delta_{ij,a}=V\langle c_{i\uparrow  a}c_{j\downarrow a}\rangle$ is
the complex order parameter. It is believed that the interlayer
tunneling processes in cuprates are mediated, to leading order, by the
spherically symmetric Cu $4s$ orbitals \cite{andersen1995lda},
although other orbitals such as those belonging to Bi and Sr atoms,
probably also play an important role. With this in mind, we adopt a
phenomenological form for the interlayer tunneling amplitude that decays exponentially with distance:
\begin{equation}\label{eq:interlayercouplingform}
	g_{ij} = g_0 e^{-(r_{ij}-c)/\rho},
      \end{equation}
where $r_{ij}^2 = c^2 + d_{ij}^2$, with $d_{ij}$ denoting the in-plane separation between two sites $i$ and $j$ on different layers and $c$ being the interlayer distance. In practice, we limit these terms to a few unit cells, beyond which the amplitudes are negligible. The characteristic radial extent of the Cu $4s$ orbital has been denoted by $\rho$. We have also considered interlayer couplings of the form $g_{ij} = g_0c^2/r_{ij}^2$ \cite{harrison2012}. Self consistent calculations still result in topological phases, but this form of coupling, when compared to \eqref{eq:interlayercouplingform} for same $g_0$, effectively corresponds to stronger interlayer coupling as the decay of $1/r^2$ is slower compared to the exponential.

\begin{figure}[t]
    \centering
    \includegraphics[width=1.\columnwidth]{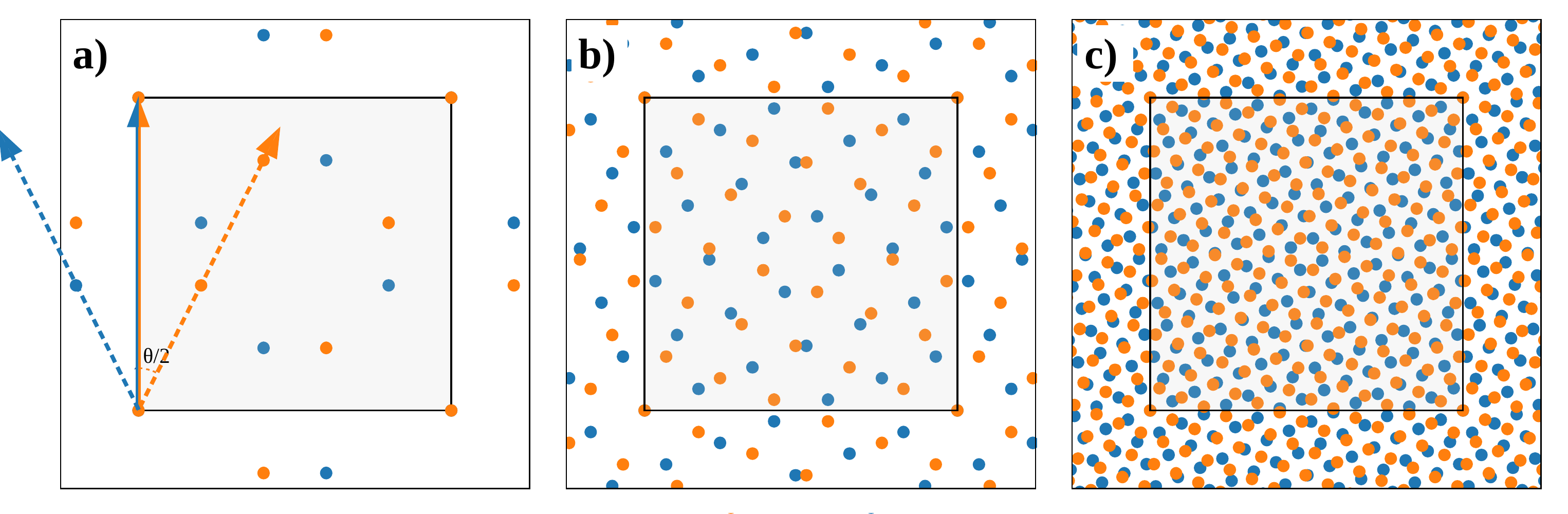}
    \caption{Geometry of the commensurate unit cells for bilayers with relative twists of  (a)  $\theta_{1,2} \simeq 53.13^{\rm o}$(b)  $\theta_{2,5} \simeq 43.60^{\rm o}$ and (c) $\theta_{5,12} \simeq 45.24^{\rm o}$. Twisting procedure is illustrated in panel (a) where orange and blue arrows indicate integer valued vectors $\bv=(m,n)$ and $\bar{\bv}=(-m,n)$ introduced in the text. Moire unit cell is shown by the shaded area. }
    \label{twists}
\end{figure}

While the above real space Hamiltonian is valid for any twist angle,
the bilayer  forms a crystal only for a discrete set of commensurate
rotation angles. To carry out the lattice mean-field calculations we
focus on commensurate twist geometries that are amenable to
numerical study using the standard Bloch representation. We note that
any commensurate twist can be described by an integer-valued `twist' vector  $\bv=(m,n)$ as shown in Fig.\ \ref{twists}. The corresponding moire unit cell comprises of $2q = 2(m^2+n^2)$ sites and the twist angle
is determined by the relation
\begin{equation}\label{theta}
 \theta_{m,n}=2\arctan{(m/n)}.   
\end{equation}
 This can be understood as a rotation of
two perfectly aligned square lattices in opposite directions
by $\theta_{m,n}/2$, which leads to the sites originally at the locations
$\bv$ and $\bar{\bv}=(-m,n)$ to lie atop each other. The unit cells for a
selection of commensurate twist angles close to $45^{\rm o}$ are
illustrated in Fig.\ \ref{twists}.

To model the Fermi surface of Bi2212 we follow
Ref.\ \cite{Klemm2001} and choose $t=153$ meV, $t'=-0.45t$ with $\mu=-1.35t$ near optimal
doping. Other parameters that determine the interlayer tunneling
amplitudes are as follows: the average spatial extent of electrons in
the $4s$ orbitals is $\rho \approx 4\text{a.u.} = 2.11 \si{\angstrom}$, the typical inter-layer distance is $c \approx 12
\si{\angstrom}$ \cite{bansil2005influence} and the in-plane separation
between the Cu atoms is $a \approx 5.4 \si{\angstrom}$ \cite{slezak2008imaging}.

In the remainder of this section we outline the procedure for obtaining gap equations for the order parameters
$\Delta_{ij,a}$ using the imaginary time path integral formalism. After
decoupling the attractive pairing potential in the Cooper channel with
the Hubbard-Stratonovich transformation, we obtain the action
corresponding to the Hamiltonian \eqref{hm_latt1} (cf. Ref.\ \cite{coleman2015introduction} for details)
\begin{equation}\label{action}
S=\sum_{\bk n} \Psi^\dag_{\bk n}[-i\omega_n + h_\bk] \Psi_{\bk n} +
\frac{\beta \cal N}{V} \sum_{\substack{i,j \in \text{u.c.} \\ a}} \Delta_{ij,a} \Delta_{ij,a}^\ast.
\end{equation}
Here $h_\bk$ is the BdG Hamiltonian that follows from  Fourier
transforming Eq.\ \eqref{hm_latt1},  $\beta = 1/k_B T$ is the inverse
temperature, $\omega_n= (2n+1)\pi/\beta$ are the Matsubara
frequencies, $V$ is the strength of attractive nearest neighbour
interactions, ${\cal N}$ is the number of unit cells and $\Psi_\bk = (\Phi_{\bk 1}, \Phi_{\bk 2})^T$ is the Nambu spinor such that $\Phi_{\bk a} = (c^{(1)}_{\bk \uparrow a}, \ldots, c^{(q)}_{\bk \uparrow a}, c^{(1)\dag}_{-\bk \downarrow a}, \ldots, c^{(q)\dag}_{-\bk \downarrow a})^{T}$, with the superscripts $1, \ldots, q$ labelling the sublattice degrees of freedom in layer $a$. 

The effective action for $\Delta_{ij,a} $ can be determined by integrating out fermionic
degrees of freedom
\begin{equation}
    S_{\text{eff}} = -\sum_{\bk n}\text{Tr}\log[-\mathcal{G}(\bk, i\omega_n )^{-1}] + \frac{\beta {\cal N}}{V} \sum_{\substack{i,j \in \text{u.c.} \\ a}} \Delta_{ij,a} \Delta_{ij,a}^\ast ,
\end{equation} 
where we defined  the Matsubara Green's function $ \mathcal{G}(\bk,
i\omega_n )= -(-i\omega_n + h_\bk)^{-1}$. To evaluate the saddle point condition for
this effective action, $\partial S_{\text{eff}}/\partial
\Delta_{ij,a}^\ast = 0$, we employ the identity $\partial_\Delta (\text{Tr}\log A)
= \text{Tr}[\partial_\Delta A A^{-1}]$. After performing the Matsubara sum, this simplifies to
\begin{equation}\label{gapeqn1}
    \Delta_{ij,a}  = -\frac{V}{{\cal N}}\sum_{\bk}\text{Tr}\left[ \frac{\partial h_\bk}{\partial \Delta_{ij,a}^\ast}  U_\bk n_F(E_\bk) U_\bk^\dag\right].
\end{equation}
In the above, $U_\bk$ is the unitary matrix that diagonalizes the Bloch
Hamiltonian as $U_\bk^\dag h_\bk U_\bk = E_\bk$ and $n_F(E_\bk)$ is a
diagonal matrix where Fermi function is applied to the eigenvalues $E_\bk$. For
decoupled layers ($g_0=0$), with the tight binding parameters given above, a self-consistent solution
converges to a $d$-wave form with maximal gap $\Delta_{\rm max} = 40$
meV, when $V=146$ meV. We use this value for calculations
where two layers are coupled ($g_0\neq 0$).  Fig.\ \ref{fig2} in the main text summarizes the results obtained from these calculations.



\subsection{Topological phases}

Here, we discuss in greater detail the various phases found
in the self-consistent treatment of the lattice model. 


To reliably determine the topology of the band structure, we turn to the Chern number, which is defined as an integral of the Berry curvature over the full Brillouin zone: $C = \frac{1}{ 2 \pi} \int_{\rm BZ} \Omega(\bk) ~ d^2k$. Computation of the Berry curvature directly from the eigenstates is complicated by the arbitrary phases that accompany numerical diagonalization. To counter this, we rely on a gauge independent formulation where \cite{bernevig2013topological}  
\begin{equation}\label{chern2}
	\Omega(\bk) = \sum_{m\neq n} 
    \frac{\langle n | \nabla_\bk h_\bk | m \rangle \times \langle m | \nabla_\bk h_\bk | n \rangle}
    {(E_m - E_n)^2}.
\end{equation}
Here $\ket{n}$ and $\ket{m}$ are eigenstates of the Bloch Hamiltonian with energies $E_n$ and $E_m$ respectively.

\begin{figure}[t]
    \centering
    \includegraphics[width=.8\columnwidth]{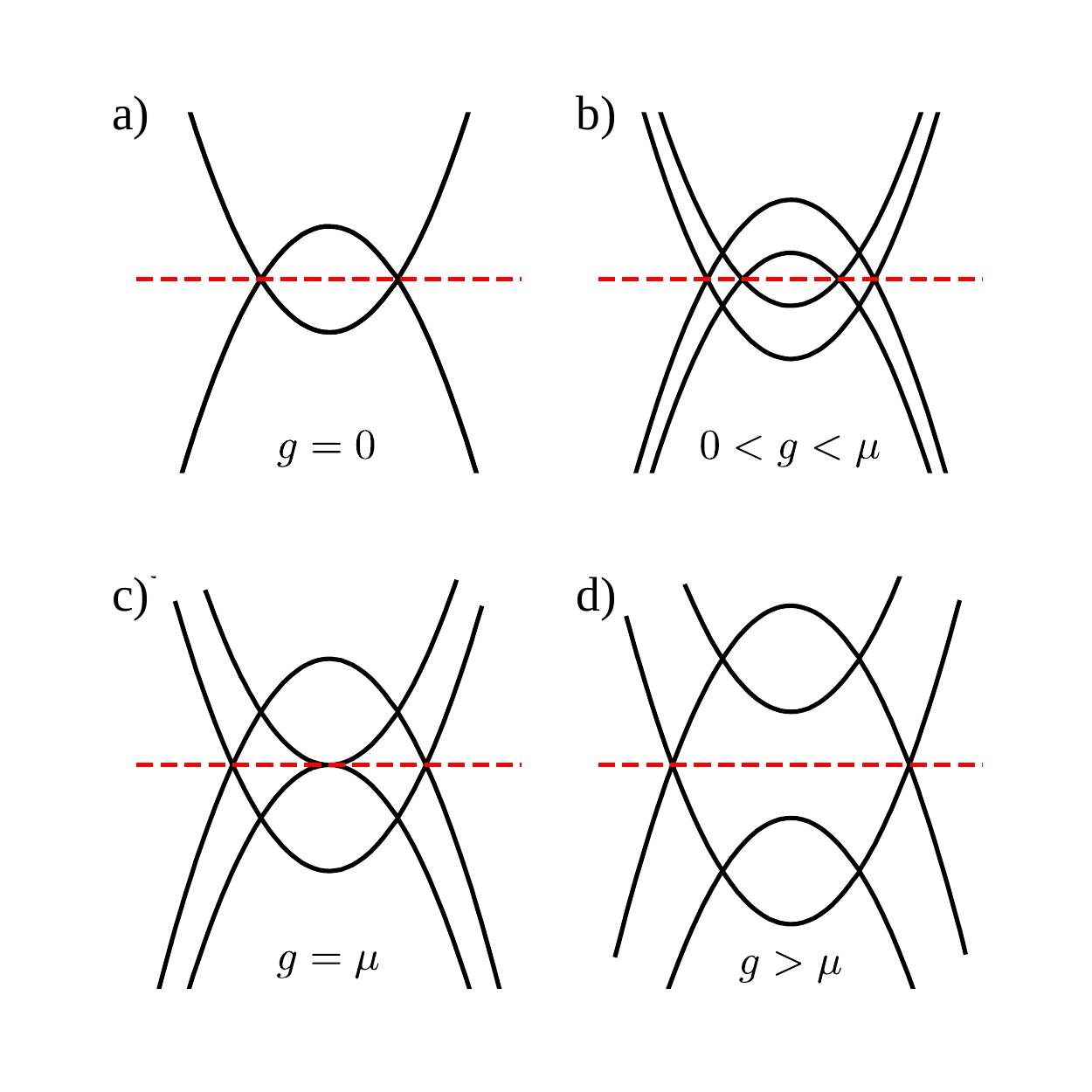}
    \caption{Schematic picture for the normal state band structure of
      the continuum model defined in
      Eq.\ \eqref{eq:continuumhamiltonian}. Both electron and hole
      bands that enter the BdG Hamiltonian are shown while the Fermi
      level is indicated by a dashed red line. Panel (a) shows
      decoupled layers with doubly degenerate bands yielding a
      circular Fermi surface. In the presence of a $d$-wave order
      parameter we have eight Dirac cones. (b) Interlayer coupling $g$
      splits the bands and  Fermi surface now consists of a pair of
      concentric circles. (c) The inner Fermi circle shrinks to a point and half of the Dirac cones disappear. (d) Four Dirac cones survive in the strong interlayer coupling ($g>\mu$) case.}
    \label{fig:c2phase}
\end{figure}

The $C=\pm 4$ and $\pm 2$ phases (indicated in Fig.\ \ref{fig2}(e) in red and green respectively)
can be understood by considering the continuum model with a circular
Fermi surface in the normal state, which we defined in Eq.
\eqref{eq:continuumhamiltonian}. Similar arguments hold for the more
realistic Fermi surface with hole pockets. When two layers are
decoupled ($g=0$), Fermi surfaces of each layer overlap, Fig.
\ref{fig:c2phase}(a),  and each layer hosts
four Dirac cones in the presence of a $d_{x^2-y^2}$ order
  parameter. Upon breaking of $\cT$, each Dirac point is gapped out and
can contribute a Berry flux of $\pm\pi$, thereby allowing for the maximal value
$C=\pm 4$. As we turn on the interlayer coupling, bands from each monolayer split,  Fig.
\ref{fig:c2phase}(b). If the interlayer coupling $g$ is strong enough
to push one of the bands above the Fermi level, half of the Dirac
cones disappear, leaving four Dirac cones behind.  In this case, the
maximum Chern number is $\pm 2$, a situation realized in
the phase diagram Fig.\ \ref{fig2}(e) for large $g_0$.

\begin{figure}[t]
    \centering
    \includegraphics[width=.7\columnwidth]{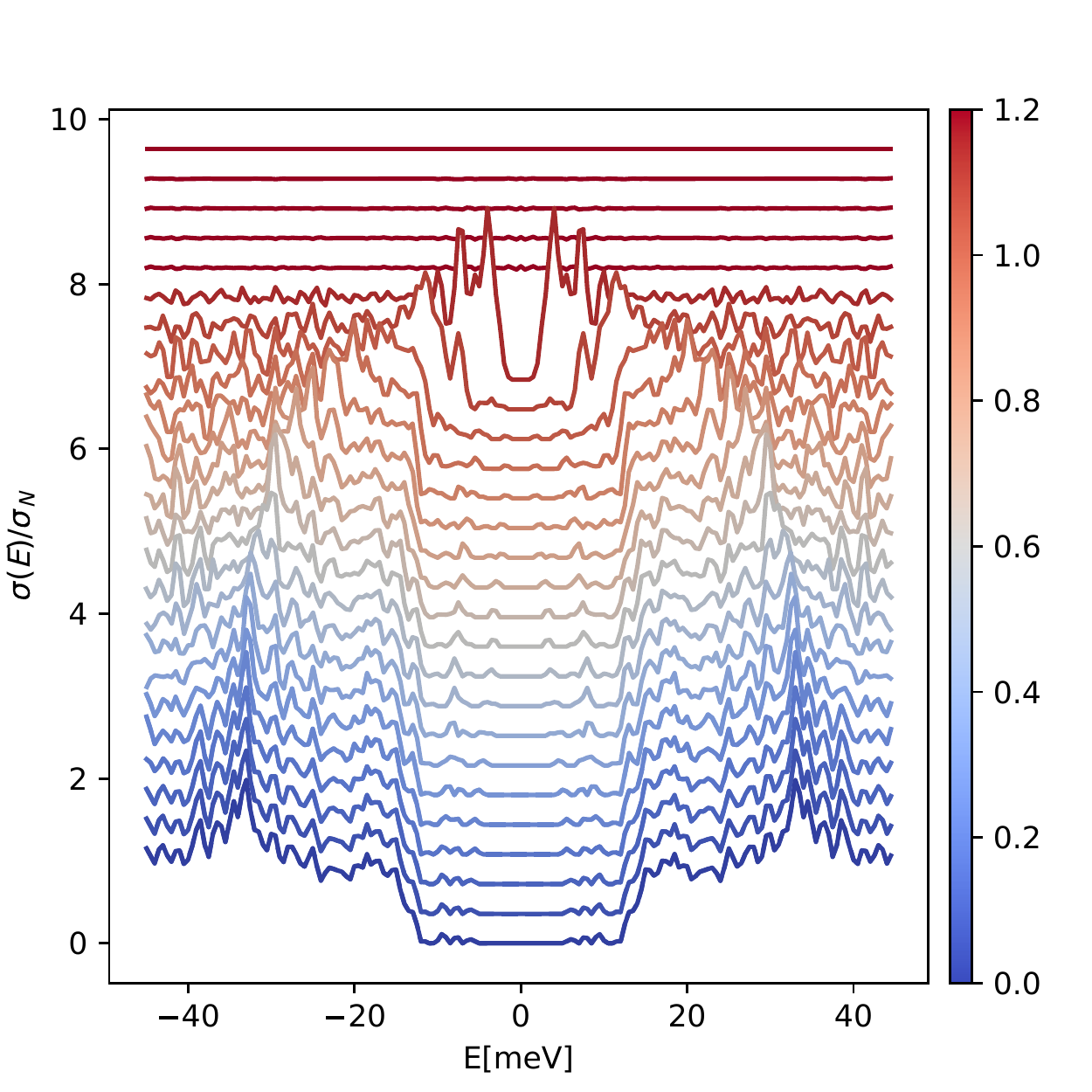}
    \caption{Evolution of the density of states with temperature in the $C=2$ phase with parameters $g_0=40$ meV and $\mu=-1.35t$. Temperature $T/T_c$ shown as color scale. The superconducting gap persists up until $T=T_c$.}
    \label{fig:dos-25}
\end{figure}

It is interesting to note that the spectral gap tends to be larger in the 
$C=2$ phase. We attribute this to the fact that the $C=2$ phase is stabilized at stronger values of the interlayer tunneling $g_0$ which  enhances the  Cooper pair tunneling amplitude. Moreover, in second order perturbation theory, the gap is expected to scale as $g_0^2/\Delta_0$. 
Temperature evolution of the corresponding tunneling density of states is shown in Fig.\ \ref{fig:dos-25}.

\begin{figure}[t]
    \centering
    \includegraphics[width=.8\columnwidth]{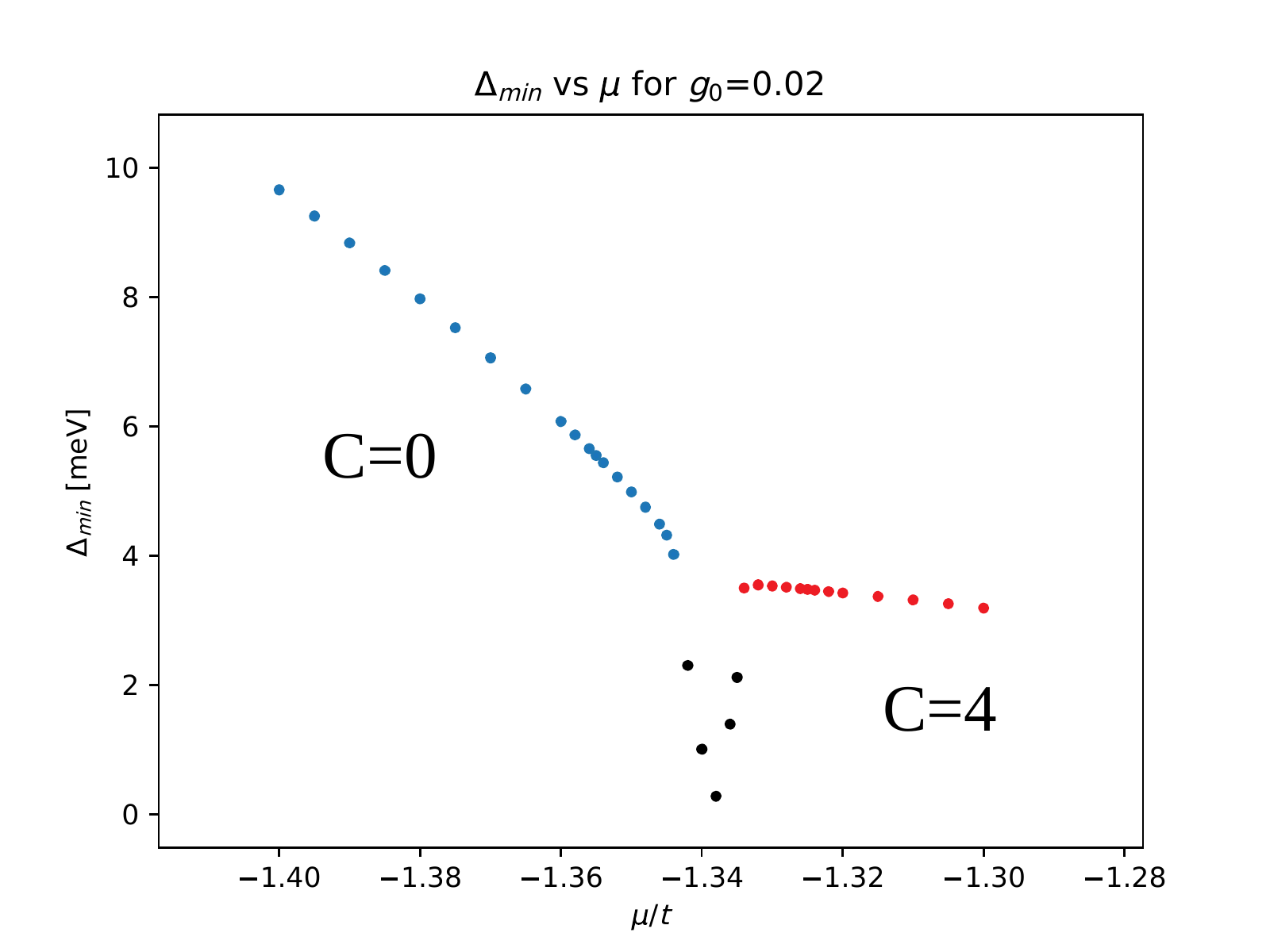}
    \caption{The topological phase transition between $C=0$ ($\mu<-1.34t$) and $C=4$ ($\mu>-1.34t$) phases is marked by a closing of the gap as the chemical potential is tuned while keeping the interlayer coupling fixed at $g_0=20$ meV and $\theta_{2,5}\simeq 43.6^{\circ}$}
    \label{fig:gapvsmuclosing}
\end{figure}

The gapped, topologically trivial $C=0$ phase in the phase diagram can be understood as follows.
As a representative of this phase, we consider the point $g_0=20$ meV and
$\mu/t=-1.3$ for $\bv=(1,3) $ in the phase diagram
Fig. \ref{fig2}(c). A self-consistent solution of the gap equation
\eqref{gapeqn1} yields the following forms for order parameters on the
two layers
\begin{align}
    \Delta_{\bk 1}=&\Delta_0(\cos{k_x}-e^{i\phi}\cos{k_y}) \\
    \Delta_{\bk 2}=&\Delta_0(\cos{k_y}-e^{i\phi}\cos{k_x})
\end{align} 
where $\Delta_1$ and $\Delta_2$ correspond to top and bottom layers respectively. The numerically obtained  phase $\phi$ tends to be close to $-\pi/4$. If we define $\Delta^d=\Delta_0e^{i\phi/2}\cos{\phi/2}$ and $\Delta^s=\Delta_0e^{i\phi/2}\sin{\phi/2}$, order parameters can be cast as
\begin{align*}
     \Delta_{\bk 1} = &\Delta^d(\cos{k_x}-\cos{k_y})  - i\Delta^s(\cos{k_x}+\cos{k_y})  \\
     \Delta_{\bk 2}=&-\Delta^d(\cos{k_x}-\cos{k_y}) - i\Delta^s(\cos{k_x}+\cos{k_y}) ,
\end{align*}
which shows that both layers have $d+is$ type superconducting order
parameter. The transition between the $d+is$ and the $d+id$ phases can be seen by tuning the chemical potential $\mu$ for a fixed interlayer coupling $g_0$, as shown in Fig. \ref{fig:gapvsmuclosing}.

\subsection{Temperature dependence}
The self-consistent gap equations can be also solved at non-zero temperature $T$. Although the Chern number is well defined only at $T=0$ we may still compute $C$ from Eq.\ \eqref{chern2} using the Bloch eigenstates obtained at $T>0$. The resulting integer then describes the underlying mean-field band structure of the system at non-zero temperature. In this case, there will be no strictly quantized physical response (such as the thermal Hall conductance), but a nonzero $C$ indicates the existence of protected chiral edge states that can be observed experimentally. In this sense the $T>0$ phases with non-zero $C$ can be regarded as topological and their properties related are to the $T=0$ counterparts.   

\begin{figure}[t]
     \centering
     \includegraphics[width=1.\columnwidth]{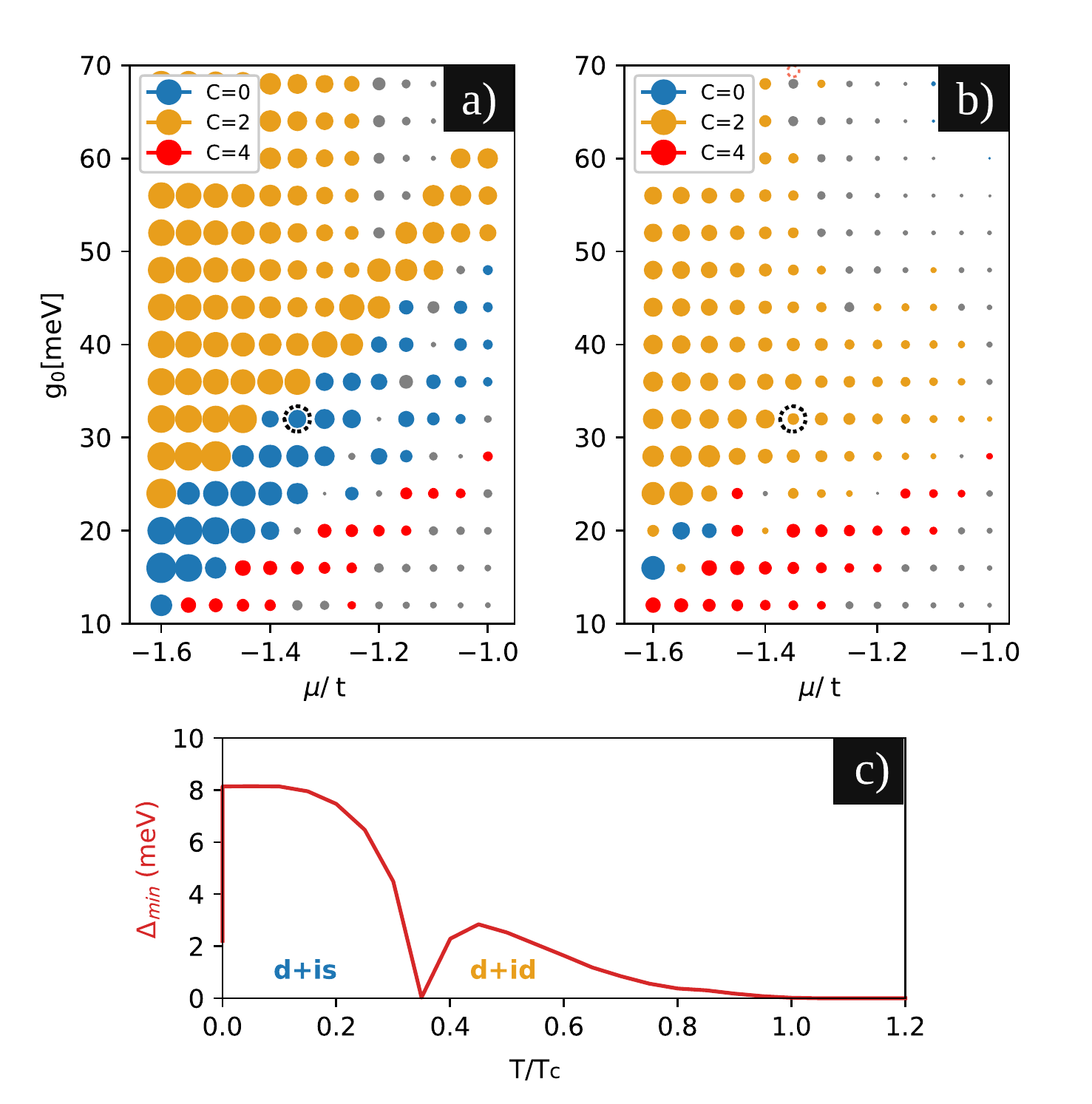}
    \caption{Temperature dependence of the phase diagram is shown for $\theta_{2,5}\simeq 43.6^{\circ}$ at $T=0$ and $T=T_c/2$ in panels (a) and (b) respectively. Panel (c) shows the minimum gap as a function of temperature for $g_0=32$meV and $\mu=-1.35t$ (the parameters are indicated in panels (a) and (b) by a dashed circle) }
    \label{fig:finiteT-phasediag}
\end{figure}

In Fig.\ \ref{fig:finiteT-phasediag} we compare phase diagrams at $T=T_c/2$ and $T=0$. We observe that, remarkably, increasing temperature tends to suppress the topologically trivial $d+is$ phase in favor of topological $d+id'$ phases. Specifically at some points in the phase diagram increasing temperature can trigger a transition from the $d+is$ phase to the $d+id'$. This is illustrated in Fig.\ \ref{fig:finiteT-phasediag}(c). As a result, we find that at elevated temperatures (but still well below $T_c$) essentially all of the phase diagram is taken over by the $C=4$ and $C=2$ topological phases.

\subsection{Multiple ${\rm CuO}_2$ planes}
\begin{figure}[t]
    \includegraphics[width=1.\columnwidth]{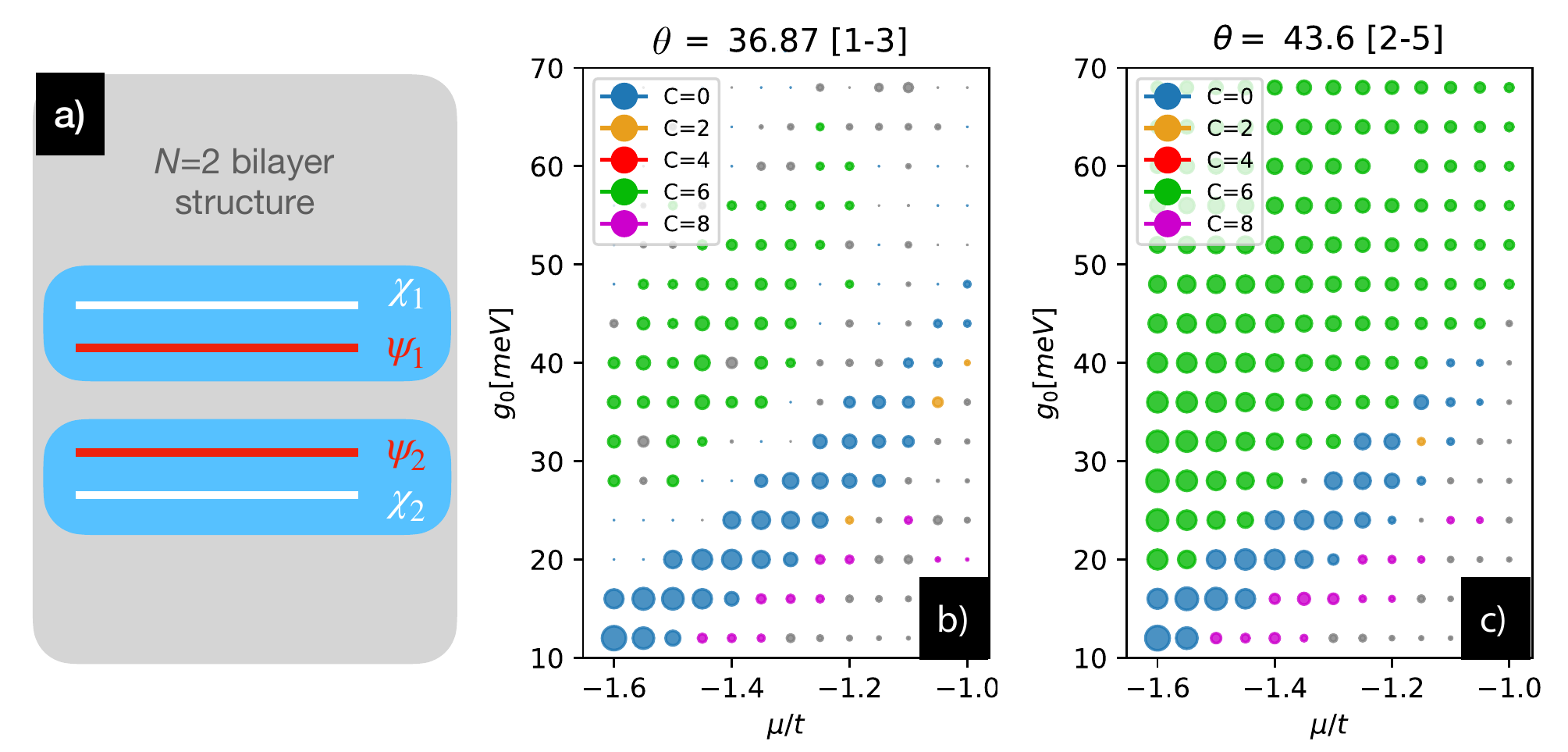}
    \caption{{\bf Physics of systems with multiple CuO$_2$ planes per monolayer.} a) The structure and notation used for the case with $N=2$ CuO$_2$ planes. Panels b) and c) show phase diagrams of the lattice BdG model for bilayers with $N=2$, relevant to the Bi2212 crystal structure with the intra-bilayer coupling set to $t_z=40$meV.}
    \label{fig4}
\end{figure}

We now briefly discuss the extension of our theory to systems with
$N>1$ CuO$_2$ planes per monolayer, focusing on compounds with $N=2$
such as Bi2212. The GL theory can be generalized by including
complex order parameters $\chi_a$ describing the extra layers as illustrated in Fig.\ \ref{fig4}(a). For $N=2$, the corresponding free energy density reads  
\begin{eqnarray}\label{f2}
 \cF&=&\cF_0
  +B(\psi_1\psi_2^\ast +{\rm c.c.})+C(\psi_1^2\psi_2^{\ast 2} +{\rm
                      c.c.}) \nonumber \\
  &+& \sum_{a=1}^2[B' (\psi_a\chi_a^\ast+{\rm  c.c.})
      +C'(\psi_a^2\chi_a^{\ast 2}+{\rm  c.c.})], 
\end{eqnarray}
where $\cF_0$ collects all the terms that are independent of the
relative phases between the layers and we only include coupling 
between adjacent layers. The most general ansatz for
order parameters which respect the requisite symmetries would be 
\begin{eqnarray}\label{f3}
  \psi_1=\psi e^{-i\varphi/2}, \ \ \psi_2=\psi e^{i\varphi/2}, \\
  \chi_1=\chi e^{-i\varphi'/2}, \ \ \chi_2=\chi e^{i\varphi'/2},\nonumber
\end{eqnarray}
with the amplitudes $\psi$ and $\chi$ real and positive. The free
energy \eqref{f2} then becomes
\begin{eqnarray}\label{f4}
\cF&=&\cF_0+2B_0\psi^2\left[-\cos(2\theta)\cos{\varphi}+
                \cK\cos({2\varphi)}\right]     \nonumber  \\
  &+&2B'_0\psi\chi\left[-\cos(\delta\varphi)+\cK'\cos{(2\delta\varphi)}\right],
\end{eqnarray}
where $\delta\varphi=(\varphi-\varphi')/2$ and  $\cK'=C'\psi\chi/B_0'$. Observe that terms describing interlayer and intralayer phase
differences, $\varphi$ and $\delta\varphi$  respectively, can be
minimized independently. The intralayer free energy,  represented by the second
line of Eq.\ \eqref{f4}, will be dominated by the single
pair tunneling term  $-\cos(\delta\varphi)$  on the account of the two
CuO$_2$  planes being in a natural untwisted configuration. This
implies $\delta\varphi=0$: 
 the two order parameters within each monolayer will be in phase, as expected. The
remaining interlayer component of the free energy then coincides with
Eq.\ \eqref{e2} and its 
analysis proceeds exactly as before. This confirms that at the
level of the GL theory, the presence of the extra layers has no effect on our prior conclusions.  

Microscopic models can likewise be extended to the $N>1$ case by
adding appropriate CuO$_2$ planes to each monolayer. We illustrate this by considering a
straightforward generalization of our lattice Hamiltonian Eq.\
\eqref{hm_latt} to $N=2$. This is achieved by adding another
square lattice of sites, which is described by the same
Hamiltonian, to each monolayer and coupling the partner lattice sites by an intralayer tunneling term
with amplitude $t_z$.  Results of the $N=2$ self-consistent calculation are
shown in Fig.\ \ref{fig4}(b,c) where we chose $t_z=40$meV as
appropriate for Bi2212 \cite{Feng2001}. We observe physics very
similar to the $N=1$ case, except that topological phases now exhibit
Chern numbers 8 and 6, consistent with the notion that in the
effective $d+id'$ state each CuO$_2$ plane contributes $C=2$ to the
aggregate Chern number of the system.

\subsection{Josephson tunneling}
In the past, several works have studied the problem of Cooper pair tunneling across $c$-axis twist junctions in order to understand the order parameter symmetry of cuprates \cite{Klemm2001, yokoyama2007theory}. Assuming a purely $d_{x^2-y^2}$ order parameter, it can be shown that the critical Josephson current must vanish when the twist angle is $45^{\circ}$. Recent experiments in twisted ultra-thin layers of Bi2212, however, observed that the critical current is essentially independent of the twist \cite{zhu2019isotropic}. Here, we show the current is indeed non-vanishing when $\cT$ is broken.

\begin{figure}[b]
    \centering
   \includegraphics[width=.9\columnwidth]{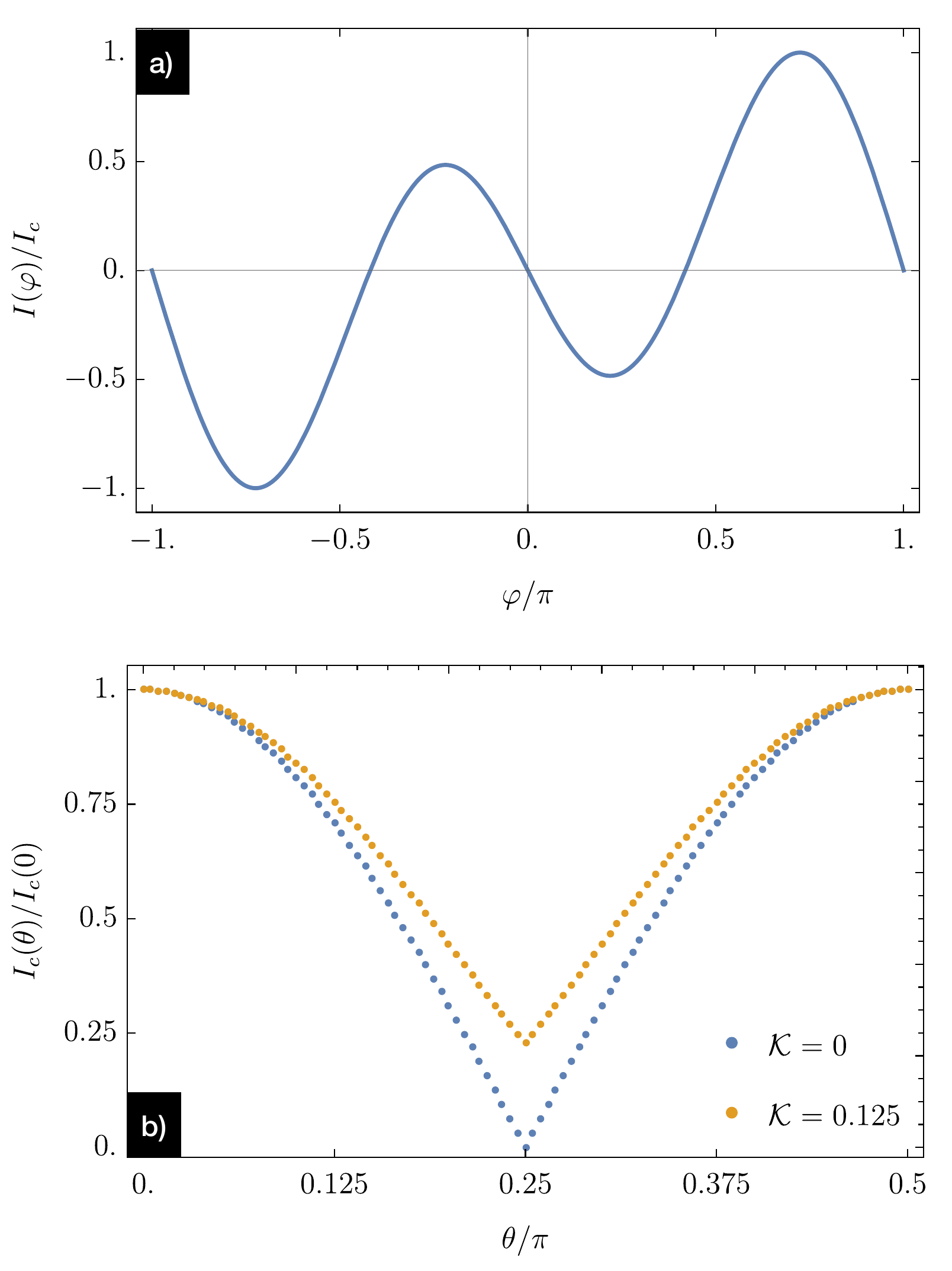}
    \caption{(a) The current-phase relation in units of the critical current at twist  $\theta = 41.4^{\circ}$ when $\cK=0.125$. 
      (b) Twist angle dependence of the critical Josephson current normalized by the current in the untwisted case. While the two curves show similar behaviour close to $\theta = 0^{\circ} \text{ and } 90^{\circ}$, a non-negative $\cK$ results in a finite value of the current at $\theta = 45^{\circ}$.}
    \label{fig:Ic}
\end{figure}

For the Josephson junction formed by the bilayer, the phase dependent supercurrent for a given $c$ axis twist can be determined within the framework of the phenomenological GL theory via the relation 
\begin{equation}
	I(\varphi, \theta) = \frac{2e}{\hbar} \frac{\partial \cF(\varphi, \theta) }{ \partial \varphi},
    \label{eq:I_phi}
\end{equation}
where the free energy has the form given in Eq.\ \eqref{e4}. A typical current-phase relation thus obtained is plotted in Fig.\ \ref{fig:Ic}(a). The critical Josephson current across the two layers for a given twist configuration, $I_c(\theta)$, may now be obtained by maximizing \eqref{eq:I_phi} with respect to $\varphi$. 

As discussed in the main text, for twists in the vicinity of $45^{\circ}$, a $\cK > 0$ ushers in a nonzero phase difference between the order parameters in the static case. This non-trivial character is manifest also in the behaviour of $I_c(\theta)$, as shown in Fig.\ \ref{fig:Ic}(b). To elucidate this, we note that setting $\cK=0$ reproduces the behaviour of a junction formed by superconductors with pure $d$-wave character \cite{Klemm2001, yokoyama2007theory}.  Note that $I_c(45^{\circ}) = 0$. For a positive $\cK$, however, the critical current at $45^{\circ}$ is non-vanishing. Given that rotations by $\theta$ and $\pi/2 - \theta$ are equivalent from the perspective of the bilayer, the functional form of the current respects this symmetry and so does its global maximum. The cusps at $\theta = 45^{\circ}$ can be seen as a consequence of this. 
%

\subsection{Density functional theory}
We solved for the total energy of the twisted bilayer unit in DFT using the meta-GGA SCAN exchange correlation potential \cite{Sun_2015}, as implemented in VASP
\cite{Kresse_1993,Kresse_1996}. Projector-Augmented-Wave pseudopotentials were
used for the calculations. Results were converged for a plane-wave
energy cutoff of 650 eV on a (5$\times$5$\times$1) mesh. Similar
results to those plotted in Fig.\ \ref{fig3} of the main text were
found in the generalized-gradient approximation \cite{Perdew1996},
with modestly a increased optimal interlayer spacing. For the GGA calculations, we added  Grimme's semi-empirical
DFT-D2 approximation \cite{Grimme2006} to better account for the long-range van der Waals binding between the bilayers.

\begin{figure}
\includegraphics[width=1.\columnwidth]{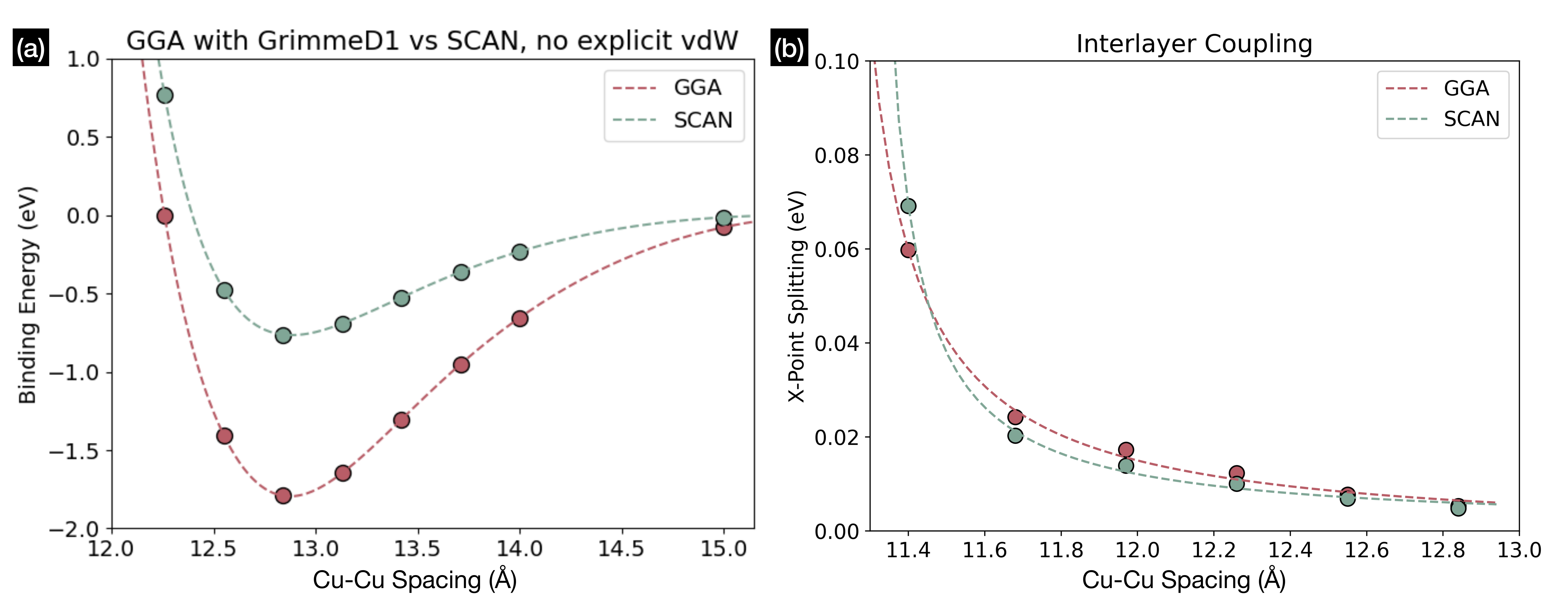}
\caption{Bilayer in Density Functional Theory: In panel (a), comparison of cohesive energy of twisted bilayer Bi2201, as computed in both the GGA and SCAN meta-GGA. Both exchange correlation potentials recover the same equilibrium spacing. (b) shows the $X$-point splitting, computed as in LDA described in the main text.}
\label{fig:SDFT}
\end{figure}

In the main text, we made an effort to parameterize the interlayer coupling by evaluating the strength of hybridization between Cu $3d$ bands at the unfolded zone edge, or $X$ point. This momentum point was chosen as previous studies of interlayer coupling in cuprates have indicated its vanishing strength along the $\Gamma M$ line \cite{Markiewicz2005}. The $X$ point is then a suitable high-symmetry point where an interlayer-coupling driven hybridization may be studied. Despite the high density of bands, we have identified a suitable band splitting which is stable and monotonic over a sufficiently wide range of interlayer spacings, so that an approximate interlayer coupling evolution can be extracted. The results, for the unshifted-unit cell, are summarized in Fig. \ref{fig3}(d) of the main text, and reproduced in Fig. \ref{fig:SDFT}(b).

The recently developed SCAN meta-GGA presents the opportunity to perform DFT on this system without introducing the empirical vdW corrections required for standard GGA. It was developed with the intention of modelling mixed-bonding systems \cite{Sun_2015}. SCAN has been successfully applied to calculations of both LSCO \cite{Furness_2018} and YBCO \cite{Zhang_2020}. To confirm the reliability of our results, and their independence of the exchange-correlation potential, we compare the results of our interlayer coupling calculations, as computed using standard GGA (with the Grimme correction), against the SCAN meta-GGA. For simplicity, we use the unshifted unit cell of Fig. \ref{fig3}(b). A summary of the cohesive energy, as well as the interlayer coupling, is provided in Fig. \ref{fig:SDFT}. Both SCAN and the vdW-corrected GGA predict a total energy minimum at 12.87 A, albeit with different cohesive energies. 
The similarities are even closer when comparing the X-point splitting, presented in Fig. \ref{fig:SDFT}(b). The splitting decays inversely with the Cu-Cu spacing, and the equilibrium splitting are nearly indistinguishable. The agreement between these results indicates the 
reliability of our $X$-point metric as an indicator of the interlayer coupling. 





\end{document}